\pdfoutput=1
\documentclass[
preprint,
titlepage,
nofootinbib,
amsmath,amssymb,
showkeys,
aps,
prd,
floatfix,
]{revtex4-1}

\usepackage{graphicx}
\usepackage{hyperref}
\usepackage[utf8]{inputenc}
\usepackage[english]{babel}
\usepackage{mathtools}
\usepackage{hyphenat}
\usepackage{natbib}

\usepackage{amsthm}
\newtheorem{theorem}{Theorem}
\newtheorem{definition}{Definition}
\newtheorem{proposition}{Proposition}

\begin{document}

\title{Gravitational radiation from birefringent matter dynamics}

\author{Nils Alex}
\email{nils.alex@fau.de}
\affiliation{Department Physik, Friedrich-Alexander-Universit{\"a}t Erlangen-N{\"u}rnberg,\\
Staudtstr.~7, 91058 Erlangen, Germany}

\begin{abstract}
We present the second-order gravitational dynamics for a spacetime inhabited by matter fields which feature vacuum birefringence. The derivation follows a perturbative variant of the covariant constructive gravity program, ensuring diffeomorphism invariance of gravity and causal compatibility of matter theory and gravity. A subsequent spatio-temporal split of this theory reveals the presence of unphysical artifacts, which are cured by imposing constraints on the gravitational constants, reducing their number from ten to seven. Within this sector, we derive the gravitational radiation emitted by a binary system in circular motion. The system emits massless waves which correspond to the radiation predicted by Einstein gravity, but also massive waves, which are generated only above a certain angular frequency threshold and are unknown to Einstein gravity. A gravitational-wave detector modeled as sphere of freely falling test masses shows quantitatively and qualitatively new behavior under the influence of this radiation. The result is a prediction of gravitational self-coupling from first principles, demonstrating the predictive power of covariant constructive gravity for modified gravity research, especially in the era of gravitational-wave astronomy.
\end{abstract}

\keywords{alternative theories of gravity, gravitational waves, diffeomorphism invariance, causality}

\maketitle

\section{Introduction}
The first earth-bound detections of gravitational waves by the LIGO and Virgo collaborations \cite{Abbott_2016,Abbott_2016_2,Abbott_2019} opened up a new avenue for research on modified gravity \cite{Yunes_2016}. These high-precision experiments demonstrated the feasibility of measuring tiny oscillations of spacetime that have their origin in faraway astrophysical events. Any modified theory of gravity that introduces changes to the generation, propagation, or detection of gravitational waves is now---in principle---falsifiable in this regard \cite{Yunes_2009,Yunes_2013}.

Using the example of \emph{area metric gravity}, we demonstrate how the covariant constructive gravity program \cite{Alex_2020} can be employed to construct gravitational theories that predict quantitatively and qualitatively new effects concerning gravitational radiation. Just like general relativity provides the dynamics for the spacetime metric governing Maxwell electrodynamics and similar field theories, area metric gravity provides the dynamics for the geometry governing a birefringent generalization of Maxwell electrodynamics.

The derivation of novel effects of gravitational radiation in area metric gravity is divided into three parts: First, in Sec.~\ref{construction}, we are concerned with the construction of area metric gravity as the gravitational theory consistent with birefringent generalizations of Maxwell electrodynamics. For this purpose, we revert to previous results \cite{Alex_2020}, but shortly review the construction procedure in order to keep the article self-contained.

In Sec.~\ref{three_plus_one_split}, we perform a 3+1 split of the thus obtained gravitational field equations. These equations turn out to be too general, allowing for unphysical behavior of solutions. Consequently, we restrict the theory to a subsector with sane phenomenology.

With the newly constructed theory at hand, we then turn to the emission of gravitational waves from a binary system in the third part, Sec.\@ \ref{binary_system}. First, we solve the problem in metric general relativity in order to establish a procedure which we subsequently adapt to area metric gravity. Finally, we consider a detector for gravitational waves modeled as sphere of freely falling test masses and derive the signal induced by gravitational radiation emitted from the binary system.

\section{Perturbative construction of area metric gravity}
\label{construction}
\subsection{General linear electrodynamics}
Our considerations start out from the assumption that spacetime is filled with matter obeying the laws of general linear electrodynamics (GLED). GLED is the most general theory of electrodynamics where electric charge and magnetic flux are conserved and the superposition principle holds \cite{Obukhov_1999,Hehl_2003}. In a very specific sense, this theory is more general than Maxwell electrodynamics: While the dynamics of the electromagnetic field in Maxwell's theory are governed by a Lorentzian metric $g$,
\begin{equation}
  S_\text{Maxwell} = \int \sqrt{-g} g^{ac} g^{bd} F_{ab} F_{cd} \, \mathrm{d}^4x,
\end{equation}
where $F$ is the field-strength 2-form, the dynamics of GLED employ a higher-rank tensor field $G$,
\begin{equation} \label{action_maxwell}
  S_\text{GLED} = \int \omega_G G^{abcd} F_{ab} F_{cd} \, \mathrm{d}^4x.
\end{equation}
The tensor field $G$ is subject to the symmetries
\begin{equation} \label{action_gled}
  G^{abcd} = G^{cdab} = -G^{bacd}
\end{equation}
and $\omega_G$ is a 1-density derived from $G$. We call $G$ the \emph{area metric} and the corresponding vector bundle $F_\text{area} \subset T^4M$ with fiber dimension 21 the \emph{area metric bundle}. Of course Maxwell electrodynamics is contained within GLED by choosing
\begin{equation}\label{induced}
  G^{abcd} = g^{ac} g^{bd} - g^{ad} g^{bc} + \sqrt{-g} \epsilon^{abcd} 
\end{equation}
and
\begin{equation}
  \quad \omega_G = \frac{1}{24} \epsilon_{abcd} G^{abcd}.
\end{equation}

A distinctive feature of GLED is the causality of light rays. The wave covector $k$ of a ray subject to Maxwell electrodynamics is constrained to the quadratic surface 
\begin{equation} \label{light_cone}
  g(k,k) = 0,
\end{equation}
which is nothing other than the well-known \emph{light cone} in relativity. In GLED, however, the surface of causal wave covectors is given by the \emph{quartic} constraint
\begin{equation} \label{fresnel_surface}
  \begin{aligned}
    P(k) \vcentcolon= {} & -\frac{1}{24} \omega_G^{-2} \epsilon_{mnpq} \epsilon_{rstu} G^{mnra} G^{bpsc} G^{dqtu} k_a k_b k_c k_d \\
                    = {} & 0.
  \end{aligned}
\end{equation}
The object $P$, often referred to as \emph{Fresnel polynomial} in the literature, is the \emph{principal polynomial} for the GLED field equations and has been calculated in this context by Rubilar \cite{Obukhov_2000}. An important qualitative difference between (\ref{light_cone}) and (\ref{fresnel_surface}) is that, in Maxwell electrodynamics, there is only \emph{one} admissible wave covector in spacetime for each spatial codirection, while in GLED there are, in general, \emph{two}.\footnote{This stems from the fact that for fixed spatial components, the constraints on the wave covector reduce to quadratic (Maxwell) or quartic (GLED) equations for the temporal component. Consequently, there are two solutions in Maxwell electrodynamics---one future-directed and one past-directed---but \emph{four} solutions in GLED, two of which are future-directed.}
The consequence is a polarization-dependent speed of light, or, more succinctly, \emph{vacuum birefringence}. In the following, we explore the gravitational ramifications of allowing for such birefringence in electrodynamics.

\subsection{Perturbative construction}
\label{constructive_gravity}
Our method of choice for deriving gravitational dynamics compatible with GLED is \emph{covariant constructive gravity} as introduced in \cite{Alex_2020}. This approach provides a precise procedure for constructing the second-order Lagrangian
\begin{equation} \label{def_lagrangian}
  \mathcal L \colon J^2F_\text{area} \rightarrow \Lambda^4M
\end{equation}
over the second-order jet bundle $J^2F_\text{area}$ based on two fundamental axioms on the dynamics from $\mathcal L$: diffeomorphism invariance and causal compatibility with matter dynamics. The perturbative variant yields a perturbative expansion of $\mathcal L$ around a flat expansion point $N$. Since our ultimate goal is the prediction of a second-order effect, we construct the Lagrangian up to third order,
\begin{equation} \label{expansion}
  \begin{aligned}
    L  = {} & a_0 + a_A H^A + a_A^{\hphantom AI} H^A_{\hphantom AI} \\
        & + a_{AB} H^A H^B + a_{AB}^{\hphantom{AB}I} H^A H^B_{\hphantom BI} + a_{A\hphantom pB}^{\hphantom Ap\hphantom Bq} H^A_{\hphantom Ap} H^B_{\hphantom Bq} \\
        & + a_{ABC} H^{A} H^{B} H^{C} + a_{ABC}^{\hphantom{ABC}I} H^{A} H^{B} H^C_{\hphantom CI} + a_{AB\hphantom pC}^{\hphantom{AB}p\hphantom{C}q} H^{A} H^B_{\hphantom Bp} H^C_{\hphantom Cq} + \mathcal{O}(H^3).
  \end{aligned}
\end{equation}
The notation is borrowed from \cite{Alex_2020}: We make use of the coordinate chart $(x^m, G^A, G^A_{\hphantom Ap}, G^A_{\hphantom AI})$ on $J^2F_\text{area}$, the coordinate representation $\mathcal L = L \mathrm d^4x$, and the coordinate deviation $H$ from the expansion point $N$,
\begin{equation}
  (H^A, H^A_{\hphantom Ap}, H^A_{\hphantom AI}) \vcentcolon= (G^A - N^A, G^A_{\hphantom Ap}, G^A_{\hphantom AI}).
\end{equation}
An appropriate expansion point is\footnote{This definition of $N$ is formulated using a coordinate-induced chart on $T^4M$. The transition to the chart on $F_\text{area}$ can be made using the intertwiner technique \cite{Alex_2020}.}
\begin{equation} \label{expansion-point}
  N^{abcd} = \eta^{a c} \eta^{b d} - \eta^{a d} \eta^{b c} + \epsilon^{a b c d},
\end{equation}
since the two requirements formulated in \cite{Alex_2020} are satisfied: $N$ is Lorentz invariant and reduces GLED to Maxwell electrodynamics on Minkowski spacetime. Hence, $N$ provides a suitable background for predicting first- and second-order gravitational effects of birefringence.

The fact that the expansion (\ref{expansion}) is around a Lorentz invariant point already reduces the coefficients $a_0,a_A,\dots$ to Lorentz invariant tensors \cite{Alex_2020}. For exactly this reason, we refrained from introducing coefficients with only a single derivative index, such as $a_A^{\hphantom Ap}$, because they drop out anyway when implementing Lorentz invariance. We will also set the coefficients $a_A$ to zero, because otherwise the flat expansion point $N$ would not be a solution to the Euler-Lagrange equations, contradicting the premise of perturbation theory. Since it is very straightforward to infer $a_0 = 0$ from diffeomorphism invariance, we also drop this coefficient.

Efficient computer algebra \cite{Reinhart_2019_sparse-tensor,Alex_2020_safe-tensor} yields a 237-dimensional basis for the remaining coefficients
\begin{equation}
  a_A^{\hphantom AI}, a_{AB}, a_{AB}^{\hphantom{AB}I}, a_{A\hphantom pB}^{\hphantom Ap\hphantom Bq}, a_{ABC}, a_{ABC}^{\hphantom{ABC}I}, a_{AB\hphantom pC}^{\hphantom{AB}p\hphantom{C}q},
\end{equation}
which is enumerated in Appendix \ref{ansaetze}. We used the same software suite \cite{Alex_2020_safe-tensor,Alex_2020_area-metric-gravity} in order to evaluate the perturbative expansion of the diffeomorphism equivariance conditions for (\ref{expansion}), which results in a linear system constraining the 237 expansion coefficients. Solving this system reduces the number of free parameters, which play the role of \emph{gravitational constants} for area metric gravity, to 50. The reduction is displayed in Appendix \ref{reduction}.

The last step of the construction procedure is to adapt the causality of the newly constructed gravitational theory to the causality of GLED. Because we constructed the theory up to second-order equations of motion, the principal polynomial is of first order. Axiom 2 formulated in \cite{Alex_2020} now requires that the corresponding null surfaces and hyperbolicity cones of the gravitational polynomial match the null surfaces and hyperbolicity cones of the GLED polynomial up to first order. To this end, we expand the polynomial (\ref{fresnel_surface}) as
\begin{equation} \label{fresnel_expansion}
  \begin{aligned}
    P_\text{GLED} =\hphantom{\vcentcolon} {} & \{\lbrack 1 - \frac{1}{24} \epsilon(H) \rbrack \eta(k,k) + \frac{1}{2} H(k,k)\}^2 + \mathcal O(H^2) \\
                  =\vcentcolon {} & \lbrack P^{(1)} \rbrack^2 + \mathcal O(H^2),
  \end{aligned}
\end{equation}
where $\epsilon(H) = \epsilon_{abcd} H^{abcd}$ and $H(k,k) = \eta_{ac} H^{abcd} k_b k_d$. Below, also the abbreviation $\eta(H) = \eta_{ac} \eta_{bd} H^{abcd}$ will be used. It is now a remarkable consequence of the diffeomorphism equivariance of (\ref{expansion}) that we actually \textbf{do not need to enforce this matching} up to our desired perturbation order, because it already follows from equivariance. In the remainder of this section we establish this fact, starting with proving that the Euler-Lagrange equations to a diffeomorphism equivariant Lagrangian are a tensor density of weight 1.

\begin{proposition} \label{euler-lagrange-theorem}
  Let $F$ be a sub-bundle of some tensor bundle over the 4-dimensional spacetime manifold $M$ and $\mathcal L \vcentcolon J^2F \rightarrow \Lambda^4 M$ be a diffeomorphism equivariant Lagrangian with coordinate representation $\mathcal L = L \mathrm d^4x$ which is degenerate in the sense that the Euler-Lagrange equations
  \begin{equation} \label{euler-lagrange}
    E_A = \frac{\delta L}{\delta G^A} = L_{:A} - D_p L_{:A}^{\hphantom{:A}p} + D_p D_q L_{:A}^{\hphantom{:A} pq},
  \end{equation}
  where $D_p f = f_{:A} G^A_{\hphantom Ap} + f_{:A}^{\hphantom{:A}q} G^A_{\hphantom Apq} + f_{:A}^{\hphantom{:A}qr} G^A_{\hphantom Apqr}$, are of second derivative order, i.e., also functions on $J^2F$. Let an infinitesimal diffeomorphism on $M$ induced by a vector field $\xi$ lift to $F$ as
  \begin{equation}
    \delta_\xi G^A = C^{A\hphantom Bm}_{\hphantom A B\hphantom m n} G^B \xi^{n}_{\hphantom n,m}.
  \end{equation}
  It follows that the Euler-Lagrange equations are diffeomorphism equivariant w.r.t.~the diffeomorphism-induced action on $\Lambda^4M \otimes F^\ast$. In particular, the local representation (\ref{euler-lagrange}) exhibits the infinitesimal transformation behavior
  \begin{equation} \label{euler-lagrange-equiv}
    \delta_\xi E_A = E_A \xi^{m}_{\hphantom{m},m} - E_B C^{B\hphantom Am}_{\hphantom B A\hphantom m n} \xi^{n}_{\hphantom n,m}.
  \end{equation}
  \begin{proof}
    The equivariance of the Lagrangian implies infinitesimally
    \begin{equation} \label{equivariance}
      \begin{aligned}
        \delta_\xi L = {} & L_{,m} \xi^{m} + L_{:A} \delta_\xi G^A + L_{:A}^{\hphantom{:A}p} \delta_\xi G^A_{\hphantom Ap} + L_{:A}^{\hphantom{:A}pq} \delta_\xi G^A_{\hphantom Apq} \\
                     = {} & L \xi^{m}_{\hphantom{m},m}.
      \end{aligned}
    \end{equation}
    Expanding $E_A$ by using its definition (\ref{euler-lagrange}) and subsequently making use of (\ref{equivariance}) in the infinitesimal transformation
    \begin{equation}
        \delta_\xi E_A = E_{A:B} \delta_\xi G^B + E_{A:B}^{\hphantom{A:B}p} \delta_\xi G^B_{\hphantom Bp} + E_{A:B}^{\hphantom{A:B}pq} \delta_\xi G^B_{\hphantom Bpq}
    \end{equation}
   yields equation (\ref{euler-lagrange-equiv}).

  \end{proof}
\end{proposition}

This property immediately translates into the \emph{principal symbol} of the Euler-Lagrange equations being a tensor density of weight 1.

\begin{proposition} \label{symbol-theorem}
  Consider the same situation as in Proposition \ref{euler-lagrange-theorem}. The principal symbol of the Euler-Lagrange equations
  \begin{equation} \label{symbol}
    T_{AB} = E_{A:B}^{\hphantom{A:B}pq} k_p k_q
  \end{equation}
  for a covector $k\in T^\ast M$ exhibits the infinitesimal transformation behavior
  \begin{equation} \label{symbol-equiv}
    \delta_\xi T_{AB} = T_{AB} \xi^{m}_{\hphantom{m},m} - T_{CB} C^{C\hphantom Am}_{\hphantom C A\hphantom m n} \xi^{n}_{\hphantom n,m} - T_{AC} C^{C\hphantom Bm}_{\hphantom C B\hphantom m n} \xi^{n}_{\hphantom n,m}.
  \end{equation}
  \begin{proof}
    A covector $k \in T^\ast M$ transforms infinitesimally as
    \begin{equation} \label{covector-equiv}
      \delta_\xi k_a = k_n \xi^n_{\hphantom n,a}.
    \end{equation}
    Expanding $T_{AB}$ using its definition (\ref{symbol}) and employing (\ref{euler-lagrange-equiv}) and (\ref{covector-equiv}) in the infinitesimal transformation
    \begin{equation}
        \delta_\xi T_{AB} = T_{AB:C} \delta_\xi G^C + T_{AB:C}^{\hphantom{AB:C}p} \delta_\xi G^C_{\hphantom Cp} + T_{AB:C}^{\hphantom{AB:C}pq} \delta_\xi G^C_{\hphantom Cpq} + \frac{\partial T_{AB}}{\partial k_a} \delta_\xi k_a
    \end{equation}
    yields (\ref{symbol-equiv}).
  \end{proof}
\end{proposition}

With the principal symbol being a tensor density of weight 1, we are now in a position to prove the central result. Four our purposes, we are only interested in Lagrangians that yield principal symbols which do not depend on derivatives of the gravitational field. Otherwise, it would be impossible to reconcile the causality of gravitational dynamics with matter dynamics, where the gravitational field only contributes locally. In other words, $T_{AB}$ is a function on $F \oplus T^\ast M$. This reduces the principal symbol to
\begin{equation}
  T_{AB} = \big\lbrack L_{:A:B}^{\hphantom{:A:B}pq} + L_{:A\hphantom{pq}:B}^{\hphantom{:A}pq} - L_{:A\hphantom{p}:B}^{\hphantom{:A}p\hphantom{:B}q} \big\rbrack k_p k_q.
\end{equation}
In particular, $T_{AB}$ is symmetric. In light of this symmetry and the diffeomorphism equivariance (\ref{euler-lagrange-equiv}), it is straightforward to see that, speaking in terms of linear algebra, the four vectors
\begin{equation} \label{kernel}
  \chi^A_{(i)} = C^{A\hphantom Bp}_{\hphantom AB\hphantom pi} G^B k_p \quad \text{for} \quad i = 1\dots4
\end{equation}
span the right and left kernel of the principal symbol,
\begin{equation}
  0 = T_{AB} \chi^A_{(i)} = T_{BA} \chi^A_{(i)}.
\end{equation}
This is a consequence of the four-dimensional gauge symmetry in diffeomorphism invariant field theory. In such a situation, where the principal symbol is a square, singular matrix, the principal polynomial $P$ is given by the adjugate matrix \cite{D_ll_2018,Itin_2009}
\begin{equation} \label{poly-def}
  \begin{aligned}
    \hskip -1em Q^{A_1\dots A_4 B_1\dots B_4} = {} & \frac{\partial^4 \operatorname{det} T}{\partial T_{A_1B_1} \dots \partial T_{A_4B_4}} \\
    = {} & \epsilon^{i_1\dots i_4} \epsilon^{j_1\dots j_4} \left\lbrack\prod_{l=1}^4 \chi^{A_l}_{(i_l)} \chi^{B_l}_{(j_l)}\right\rbrack P.
  \end{aligned}
\end{equation}
In particular, $P$ is a homogeneous polynomial of degree $2N - 16$, with $N$ being the fiber dimension of $F$. We now turn back to the bundle in question, $F_\text{area}$, and prove that, up to second perturbation order in the Euler-Lagrange equations and, consequently, up to first perturbation order in the principal polynomial, there is no causality mismatch left to be fixed. Diffeomorphism invariance of the gravitational dynamics is sufficient to constrain the principal polynomial to the GLED polynomial.

\begin{theorem} \label{poly-match}
  Consider the same situation as in Proposition \ref{euler-lagrange-theorem} with $F=F_\mathrm{area}$. Let $T_{AB}$ be independent from the derivatives of the gravitational field, i.e., be a function on $F \oplus T^\ast M$. The principal polynomial $P$, as defined in (\ref{poly-def}), is a scalar density of weight 57. In particular, it exhibits the infinitesimal transformation behavior
  \begin{equation}
    \delta_\xi P = 57 \cdot P \xi^{m}_{\hphantom m,m}.
  \end{equation}
  To first order in the expansion $G = N + H$ of the area metric field with $N$ as in (\ref{expansion-point}), the principal polynomial is equivalent to the GLED principal polynomial $P^{(1)}$,
  \begin{equation} \label{poly-equivalent}
    P = \lbrack\omega P^{(1)}\rbrack^{13} + \mathcal O(H^2).
  \end{equation}
  $\omega$ denotes a $\frac{57}{13}$-density on $F_\mathrm{area}$. In particular, both polynomials describe the same null surfaces and hyperbolicity cones.
  \begin{proof}
    The area metric field transforms under infinitesimal spacetime diffeomorphisms as
    \begin{equation}\label{area-trafo}
      \delta_\xi G^A = C^{A\hphantom{B}m}_{\hphantom{A}B\hphantom{m}n} G^B \xi^{n}_{\hphantom{n},m} = -4 J^A_{\hphantom{A}abcn} I^{abcm}_{\hphantom{abcd}B} G^B \xi^{n}_{\hphantom{n},m}.
    \end{equation}
    $I$ and $J$ are a choice of constant injection and surjection, respectively, relating $T^4M$ with its sub-bundle $F_\text{area}$ such that $J \circ I = \mathrm{id}$ \cite{Alex_2020}. It is straightforward to see that the functions $\chi^A_{(i)}$ spanning the left and right kernel of $T_{AB}$ are tensor-valued, i.e., transform infinitesimally as
    \begin{equation}
      \delta_\xi \chi^A_{(i)} = C^{A\hphantom{B}m}_{\hphantom{A}B\hphantom{m}n} \chi^B_{(i)} \xi^n_{\hphantom{n},m} + \chi^A_{(m)} \xi^m_{\hphantom{m},i}.
    \end{equation}
    Putting everything together, we first calculate
    \begin{equation} \label{q-equiv}
      \begin{aligned}
        & \hskip -2em \delta_\xi Q^{A_1\dots A_4 B_1 \dots B_4} \\
        = {} & \delta_\xi \left\lbrack \frac{4}{21!} \epsilon^{A_1\dots A_{21}} \epsilon^{B_1\dots B_{21}} T_{A_5B_5} \cdots T_{A_{21}B_{21}} \right\rbrack \\
        = {} & 59 \cdot Q^{A_1\dots A_4 B_1 \dots B_4} \xi^m_{\hphantom m,m} \\
        &  + C^{A_1\hphantom{A}m}_{\hphantom{A_1}A\hphantom{m}n} Q^{AA_2A_3A_4B_1\dots B_4} \xi^n_{\hphantom n,m} + \dots \\
        & + C^{B_4\hphantom{B}m}_{\hphantom{B_4}B\hphantom{m}n} Q^{A_1\dots A_4B_1B_2B_3B} \xi^n_{\hphantom n,m}
      \end{aligned}
    \end{equation}
    and
    \begin{equation} \label{chichi-equiv}
      \begin{aligned}
        & \hskip -2em \delta_\xi \left\lbrack \epsilon^{a_1\dots a_4} \epsilon^{b_1\dots b_4} \prod_{i=1}^{4}\chi^{A_i}_{(a_i)} \chi^{B_i}_{(b_i)} \right\rbrack \\
        = {} & 2 \cdot \epsilon^{a_1\dots a_4} \epsilon^{b_1\dots b_4} \xi^{m}_{\hphantom m,m} \prod_{i=1}^{4} \chi^{A_i}_{(a_i)} \chi^{B_i}_{(b_i)} \\
        & + C^{A_1\hphantom{A}m}_{\hphantom{A_1}A\hphantom{m}n} \epsilon^{a_1\dots a_4} \epsilon^{b_1\dots b_4} \xi^{n}_{\hphantom n,m} \chi^{A}_{(a_1)} \cdots \chi^{B_4}_{(b_4)} + \cdots \\
        & +C^{B_4\hphantom{B}m}_{\hphantom{B_4}B\hphantom{m}n} \epsilon^{a_1\dots a_4} \epsilon^{b_1\dots b_4} \xi^{n}_{\hphantom n,m} \chi^{A_1}_{(a_1)} \cdots \chi^{B}_{(b_4)}.
      \end{aligned}
    \end{equation}
    When verifying both calculations, the identities $\epsilon^{[a_1\dots a_4} T^{a] \dots} = 0$ and $\epsilon^{[A_1\dots A_{21}} T^{A] \dots} = 0$ come in handy. Substituting (\ref{q-equiv}) and (\ref{chichi-equiv}) in the infinitesimal transformation of $P$ as defined in (\ref{poly-def}) yields the transformation of a density of weight 57,
    \begin{equation}
      \delta_\xi P = 57 \cdot P \xi^{m}_{\hphantom m,m}.
    \end{equation}
    This is equivalent to the symmetric coefficients $P^{a_1\dots a_{26}}$ being a tensor density of weight 57. For such a bundle function, we set up the diffeomorphism equivariance equations in PDE form (see \cite{Alex_2020})
    \begin{align}
      P^{a_1\dots a_{26}}_{\hphantom{a_1 \dots a_{26}},m} = {} & 0 \label{p-equiv-const} \\
      P^{a_1\dots a_{26}}_{\hphantom{a_1 \dots a_{26}}:A} C^{A\hphantom{B}m}_{\hphantom{A}B\hphantom{m}n} G^B = {} & 57 \cdot P^{a_1\dots a_{26}} - 26 \cdot P^{m(a_1\dots a_{25}} \delta^{a_{26})}_{\hphantom{a_{26})}n}. \nonumber
    \end{align}
    The constant Lorentz-invariant ansatz to first order reads\footnote{The first coefficient can be absorbed into an irrelevant overall factor, so we set it to 1.}
    \begin{equation}
        P = \eta(k,k)^{13} + A\cdot\epsilon(H) \eta(k,k)^{13} + B\cdot\eta(H) \eta(k,k)^{13} + C\cdot H(k,k) \eta(k,k)^{12} + \mathcal O(H^2).
    \end{equation}
    Evaluating and solving the equivariance equations results in the most general principal polynomial of area metric gravity to first order, which after ``completing the thirteenth power''\footnote{$1+\epsilon+\mathcal{O}(\epsilon^2) = \lbrack 1+\frac{\epsilon}{13} \rbrack^{13} + \mathcal{O}(\epsilon^2)$} amounts to
    \begin{equation}
        P = \left\{ \left\lbrack 1 - \frac{35}{12\cdot 13} \epsilon(H) + \frac{A}{13} \cdot \left(\eta(H) - \frac{1}{2} \epsilon(H)\right)\right\rbrack \eta(k,k) +  \frac{1}{2} H(k,k) \right\}^{13} + \mathcal O(H^2).
    \end{equation}
    Using the same procedure as above to derive the most general scalar density $\omega$ of weight $\frac{57}{13}$ on $F_\text{area}$, we find
    \begin{equation}
      \omega = 1 + A\left\lbrack \eta(H) - \frac{1}{2} \epsilon(H)\right\rbrack - \frac{19}{8\cdot 13} \epsilon(H)
    \end{equation}
    and by simple multiplication of $\omega$ with $P^{(1)}$ (see Eq.~\ref{fresnel_expansion}) finally verify assertion (\ref{poly-equivalent}).
  \end{proof}
\end{theorem}

\section{3+1 split of area metric gravity}
\label{three_plus_one_split}
\subsection{Sliced spacetime}

Because the field equations to the just devised theory are---as it will turn out---hyperbolic, we now turn to a 3+1 formulation where the initial value problem becomes manifest. This will later be the starting point for the prediction of gravitational radiation in Sec.~\ref{binary_system}.
\begin{definition}[slicing]
  Let $M$ be a four-dimensional spacetime manifold. A slicing of $M$ is a diffeomorphism
  \begin{equation}\label{slicing}
    \phi \colon \Sigma \times \mathbb R \rightarrow M,
  \end{equation}
  where $\Sigma$ is the three-dimensional spatial manifold.
\end{definition}
Note that such a diffeomorphism always exists, as we consider a matter theory with a well-defined initial value problem\footnote{The GLED principal polynomial is hyperbolic for certain algebraic classes of area metrics, in particular for the class containing the flat expansion point $N$ \cite{Schuller_2010}.}, mandating the existence of a spatial manifold $\Sigma$ for the prescription of initial data. The slicing (\ref{slicing}) is not unique: Any diffeomorphism $\psi\colon M \rightarrow M$ yields another slicing $\tilde \phi = \psi \circ \phi$.

With every slicing comes a holonomic basis
\begin{equation}
  \frac{\partial}{\partial x^a} = \left(\frac{\partial}{\partial t}, \frac{\partial}{\partial x^\alpha}\right)
\end{equation}
of the tangent spaces $T_{\phi(s,\lambda)}M$, constructed as pushforwards of holonomic bases on $T_s\Sigma$ and $T_\lambda\mathbb R$. In the same fashion, a holonomic basis
\begin{equation}
  \mathrm dx^a = (\mathrm dt, \mathrm dx^\alpha)
\end{equation}
of co-tangent spaces follows from the slicing. This split of $TM$ and $T^\ast M$ carries over to higher-rank tensor bundles, proper sub-bundles thereof and corresponding jet bundles, including $J^2F_\text{area}$.

We define spatial quantities using an observer definition for arbitrary tensor theories \cite{Giesel_2012}. This definition only makes use of the principal polynomial. An \emph{observer frame} is a nonholonomic frame $(T, e_\alpha = \frac{\partial}{\partial x^\alpha})$ together with a dual co-frame $(n = \lambda\cdot\mathrm dt, \epsilon^\alpha)$, subject to the conditions
\begin{equation}\label{frame-conditions}
  P(n) = 1 \quad \text{and} \quad T = \frac{1}{\operatorname{deg} P} \frac{DP(n)}{P(n)}.
\end{equation}
We decompose the time direction using the observer frame into \emph{lapse} $N$ and \emph{shift} $N^\alpha$
\begin{equation}
  \frac{\partial}{\partial t} = N T + N^{\alpha} \frac{\partial}{\partial x^\alpha}
\end{equation}
and perform the spatio-temporal split of $F_\text{area}$ in terms of observer quantities (see also \cite{Giesel_2012})
\begin{equation}
\begin{aligned}
  G(\mathrm dt,\mathrm dx^\alpha,\mathrm dt,\mathrm dx^\beta) = {} & \frac{1}{N^2} G(n,\epsilon^\alpha,n,\epsilon^\beta), \\
    G(\mathrm dt,\mathrm dx^\alpha,\mathrm dx^\beta,\mathrm dx^\gamma) = {} & - \frac{2}{N^2} G(n,\epsilon^\alpha,n,\epsilon^{\lbrack\gamma}) N^{\beta\rbrack} + \frac{1}{N} G(n,\epsilon^\alpha,\epsilon^\beta,\epsilon^\gamma), \\
    G(\mathrm dx^\alpha,\mathrm dx^\beta,\mathrm dx^\gamma,\mathrm dx^\delta) = {}  & \frac{4}{N^2} N^{\lbrack\alpha} G(n,\epsilon^{\beta\rbrack},n,\epsilon^{\lbrack\delta}) N^{\gamma\rbrack} \\
                              & + \frac{2}{N} N^{\lbrack\alpha} G(n,\epsilon^{\beta\rbrack},\epsilon^\gamma,\epsilon^\delta) + \frac{2}{N} N^{\lbrack\gamma} G(n,\epsilon^{\delta\rbrack},\epsilon^\alpha,\epsilon^\beta) \\
                              & + G(\epsilon^\alpha,\epsilon^\beta,\epsilon^\gamma,\epsilon^\delta).
\end{aligned}
\end{equation}
It is convenient to introduce the fields
\begin{equation}
\begin{aligned}
    \hat G^{\alpha \beta} = {} & -G(n,\epsilon^\alpha,n,\epsilon^\beta), \\
    \hat G^{\alpha}_{\hphantom \alpha\beta} = {} & \frac{1}{2} (\omega_{\hat G})^{-1} \epsilon_{\beta\mu\nu} G(n,\epsilon^\alpha,\epsilon^\mu,\epsilon^\nu) - \delta^{\alpha}_{\hphantom\alpha \beta}, \\
    \hat G_{\alpha \beta} = {} & \frac{1}{4} (\omega_{\hat G})^{-2} \epsilon_{\alpha\mu\nu} \epsilon_{\beta\rho\sigma} G(\epsilon^\mu,\epsilon^\nu,\epsilon^\rho,\epsilon^\sigma),
\end{aligned}
\end{equation}
with
\begin{equation}
  \omega_{\hat G} = \sqrt{\operatorname{det} \hat G^{\cdot\cdot}}.
\end{equation}
Obviously, $\hat G^{\alpha \beta}$ and $\hat G_{\alpha \beta}$ are symmetric. Moreover, it follows from the frame conditions (\ref{frame-conditions}) for the GLED polynomial (\ref{fresnel_surface}) that $\hat G^{\alpha}_{\hphantom \alpha\beta}$ is symmetric w.r.t.~$\hat G^{\alpha \beta}$ and trace-free. We thus have a decomposition of the 21 spacetime components of $G$ into 17 observer quantities $\hat G$, 3 shift components $N^\alpha$, and the lapse $N$---similar to the 3+1 decomposition of a spacetime metric $g$ into shift, lapse, and a spatial metric $\hat g$.

For the perturbative formulation of area metric gravity, we expand the observer quantities around the flat expansion point $N$ (\ref{expansion-point}) as
\begin{equation}\label{quantities1}
  \begin{aligned}
    N = {} & 1 + A, \\
    N^\alpha = {} & b^\alpha, \\
    \hat G^{\alpha \beta} = {} & \gamma^{\alpha \beta} + h^{\alpha \beta}, \\
    \hat G^{\alpha}_{\hphantom \alpha\beta} = {} & k^{\alpha}_{\hphantom \alpha\beta}, \\
    \hat G_{\alpha\beta} = {} & \gamma_{\alpha \beta} + l_{\alpha\beta}.
  \end{aligned}
\end{equation}
From now on, spatial indices will be raised and lowered at will using the flat spatial metric $\gamma$ and its inverse. Instead of working with the perturbations $h,k,l$ directly, we define a more convenient set of fields which will later on decouple in the field equations,
\begin{equation}\label{quantities2}
  u^{\alpha \beta} = h^{\alpha\beta} - l^{\alpha\beta},\quad v^{\alpha \beta} = h^{\alpha\beta} + l^{\alpha\beta},\quad w^{\alpha\beta} = 2 k^{\alpha\beta}.
\end{equation}

\subsection{Gauge fixing}
\label{gauge_fixing}

Before we present the gravitational field equations in terms of these fields, we will fix the gauge symmetry we deliberately introduced by making the theory diffeomorphism invariant. To this end, we employ Helmholtz's theorem and decompose the shift perturbation into a longitudinal scalar $B$ and a transverse vector $B^\alpha$ with $\partial_\alpha B^\alpha = 0$,
\begin{equation}
  b^\alpha = \partial^\alpha B + B^\alpha.
\end{equation}
On the same basis, we decompose the field $u^{\alpha\beta}$ into two scalars $U$ and $\tilde U$, a transverse vector $U^\alpha$ with $\partial_\alpha U^\alpha = 0$, and a transverse traceless tensor $U^{\alpha\beta}$ with $\partial_\alpha U^{\alpha\beta} = 0$ and $\gamma_{\alpha\beta} U^{\alpha\beta} = 0$,
\begin{equation}
  u^{\alpha\beta} = U^{\alpha\beta} + 2 \partial^{(\alpha} U^{\beta)} + \gamma^{\alpha\beta} \tilde U + \Delta^{\alpha\beta} U,
\end{equation}
where the scalar $U$ enters via the traceless Hessian $\Delta_{\alpha\beta} = \partial_\alpha\partial_\beta - \frac{1}{3}\gamma_{\alpha\beta}\Delta$. The fields $v^{\alpha\beta}$ and $w^{\alpha\beta}$ decompose in a similar way, but with $w^{\alpha\beta}$ being traceless, there is no scalar $\tilde W$.

A gauge transform is infinitesimally represented by a vector field $\xi$ (see Eq.~\ref{area-trafo}), such that the perturbation $H$ transforms as
\begin{equation}
  H^{\prime A} = H^A + C^{A\hphantom Bm}_{\hphantom AB\hphantom mn} N^B \xi^n_{\hphantom n,m}.
\end{equation}
Inspecting the individual components of $H^{\prime A}$, we notice that the four components of $\xi$ can be chosen such that the four gauge conditions
\begin{equation}\label{area-gauge}
  \begin{aligned}
    0 = {} & B, \\
    0 = {} & U^{\alpha} - V^{\alpha}, \\
    0 = {} & U + V
  \end{aligned}
\end{equation}
are satisfied (see \cite{Schneider_2017}). Adopting this choice leaves us with 17 degrees of freedom in the scalars $A, \tilde U, \tilde V, V, W$, the transverse vectors $B^\alpha, U^\alpha, W^\alpha$, and the transverse traceless tensors $U^{\alpha\beta}, V^{\alpha\beta}, W^{\alpha\beta}$.

\subsection{Field equations}
\label{field-equations}
Applying the decomposition of the area metric field into observer quantities to the Lagrangian (\ref{expansion}) and performing variations w.r.t.~the 21 degrees of freedom yields 21 field equations, four of which are redundant as a consequence of the Noether theorem for the gauge symmetry. This calculation has been carried out using the field-theory motivated computer algebra system \textsc{Cadabra} \cite{Peeters2007,Peeters_2018} and the previously computed ansätze and solutions.\footnote{The \textsc{Cadabra} code is publicly available at \cite{Alex_2020_area-metric-gravity}.} In this process, we observe that only a subset of gravitational constants appearing in the Lagrangian enter the field equations. Up to first order, the number of those constants is 10.

The gravitational field equations are displayed in their entirety in Appendix \ref{linear-eom}. In the following we will see that there are exemplary cases which show that the first-order theory still allows for unphysical phenomenology. Restricting the theory to a physical subset manifests itself in a further reduction of the first-order gravitational constants from 10 to 7.

First, consider the scalar equations for the gravitational field sourced by a point mass $M$ at rest at the origin of our chart, i.e., with world line
\begin{equation}
  \gamma^a(\lambda) = \lambda \delta^a_{\hphantom a0}.
\end{equation}
This point mass shall be an idealization of a matter field adhering to GLED dynamics. As such, its dynamics are given by the action \cite{R_tzel_2011}
\begin{equation}
  S_\text{matter}[\gamma] = -M \int \mathrm d\lambda P_\text{GLED}(L^{-1}(\dot{\gamma}(\lambda)))^{-\frac{1}{4}},
\end{equation}
with $L^{-1}$ being the inverse of the Legendre map associated to $P_\text{GLED}$.\footnote{In the Maxwell-Einstein equivalent, this action measures the length of the particle world line, which is to be maximized.} Perturbatively, we obtain the nonvanishing contribution
\begin{equation}\label{source}
  \frac{\delta S_\text{matter}}{\delta A(x)} = -M \delta^{(3)}(x).
\end{equation}
Since the matter distribution is stationary, we also consider stationary gravitational fields by assuming that all time derivatives vanish. Time-dependent fields would be solutions to the homogeneous field equations which can, of course, be added at will. The stationary scalar equations (\ref{scalar-equations-appendix}) sourced by (\ref{source}) then take the form
\begin{equation}\label{stationary-eqns}
  E^\text{(scalar)}_i = M \delta^{(3)}(x) \delta^0_i + \sum_j\left\lbrack a_{ij} S_j + b_{ij} \Delta S_j + c_{ij} \Delta\Delta S_j \right\rbrack.
\end{equation}

Solving these equations yields a mix of long-ranging Coulomb potentials $\propto \frac{1}{r}$ and short-ranging Yukawa potentials $\propto \frac{1}{r} \mathrm e^{-\mu r}$. The exact nature of this mix and the scales of the Yukawa potentials follow from the gravitational constants. This result is greatly simplified by imposing a condition on its phenomenology: The solution to (\ref{stationary-eqns}) shall be given by the linearized Schwarzschild solution of general relativity for a central mass $M$ plus only short-ranging Yukawa corrections. Using above observer frame definition, the $3+1$ decomposition of a spacetime metric $g$ into a spatial metric $\hat g$, shift, and lapse reads
\begin{equation}\label{metric-decomposition}
  \begin{aligned}
    g^{00} = {} & \frac{1}{N^2} &\approx {} & 1 - 2A, \\
    g^{0\alpha} = {} & -\frac{N^\alpha}{N^2} &\approx {} & -b^\alpha, \\
    g^{\alpha\beta} = {} & \frac{N^\alpha N^\beta}{N^2} - \hat g^{\alpha\beta} &\approx {} & -\gamma^{\alpha\beta} - \varphi^{\alpha\beta}.
  \end{aligned}
\end{equation}
Inserting this decomposition into the metrically induced area metric (\ref{induced}) yields the spatial area metric fields
\begin{equation}
  \begin{aligned}
    \hat G^{\alpha\beta} = {} & \hat g^{\alpha\beta} &= {} & \gamma^{\alpha\beta} + \varphi^{\alpha\beta}, \\
    \hat G^{\alpha}_{\hphantom\alpha\beta} = {} & 0, \\
    \hat G_{\alpha\beta} = {} & \left({\hat g}^{-1}\right)_{\alpha\beta} &\approx {} & \gamma_{\alpha\beta} - \varphi_{\alpha\beta}.
  \end{aligned}
\end{equation}
Comparing with the definitions of perturbative area metric observer quantities (\ref{quantities1}) and (\ref{quantities2}) we find the metrically induced perturbation
\begin{equation} \label{induced_area_perturbative}
  u^{\alpha \beta} = 2\varphi^{\alpha\beta},\quad v^{\alpha \beta} = 0,\quad w^{\alpha\beta} = 0.
\end{equation}
The metric solution around a stationary point mass to first order is quickly obtained by expanding the well-known Schwarzschild solution \cite{Schwarzschild_1916} to this order, which gives
\begin{equation} \label{linear_schwarzschild_metric}
  A \propto \frac{1}{r} \quad \text{and} \quad \varphi^{\alpha\beta} = 2A\gamma^{\alpha\beta}.
\end{equation}
With (\ref{induced_area_perturbative}) and (\ref{linear_schwarzschild_metric}) in mind, the condition that the stationary scalar equations (\ref{stationary-eqns}) be solved by short-ranging Yukawa corrections of the metrically induced linearized Schwarzschild solution can now be formulated as
\begin{equation}
  \begin{aligned}
    4A - \tilde U = {} & (\text{Yukawa corrections}), \\
    V = {} & (\text{Yukawa corrections}), \\
    W = {} & (\text{Yukawa corrections}), \\
    \tilde V = {} & (\text{Yukawa corrections}).
  \end{aligned}
\end{equation}
These conditions translate into two conditions on the 10 gravitational constants governing first-order area metric gravity. Incorporating both conditions, the solution to (\ref{stationary-eqns}) reads
\begin{equation}\label{scalar-solution}
  \begin{aligned}
    V(x) = {} & 0, \\
    W(x) = {} & 0, \\
    \tilde U(x) = {} & \frac{M}{4\pi r} \left\lbrack \alpha + \beta \mathrm e^{-\mu r}\right\rbrack, \\
    \tilde U(x) = {} & \frac{M}{4\pi r} \left\lbrack \gamma \mathrm e^{-\mu r}\right\rbrack, \\
    \tilde A(x) = {} & \frac{M}{4\pi r} \left\lbrack \frac{1}{4}\alpha - \frac{1}{4} (\beta + 3 \gamma) \mathrm e^{-\mu r}\right\rbrack.
  \end{aligned}
\end{equation}
$\alpha$, $\beta$, $\gamma$, and $\mu$ are 4 independent combinations of the 8 remaining gravitational constants.\footnote{See Appendix \ref{linear-eom} for the details.} From now on, we will work in this sector of the theory, which we deem the phenomenologically most relevant one.

The second unphysical phenomenon still present in the theory is a divergence in the time evolution of some modes. Inspecting, for example, the equations of motion for the transverse traceless tensor fields \emph{in vacuo} (\ref{tt-eqns-reduced}), we find coupled equations of the kind\footnote{$\Box u = \ddot u - \Delta u$}
\begin{equation}
  \begin{aligned}
    0 = {} & \Box u + \nu^2 u + \sigma v, \\
    0 = {} & \Box v + \nu^2 v - \sigma u.
  \end{aligned}
\end{equation}
Performing a spatial Fourier transform, we find the four eigenvalues for the time evolution of a mode $k$,
\begin{equation}
  \lambda_k = \pm \mathrm i \sqrt{(k^2+\nu^2) \pm \mathrm i\sigma}.
\end{equation}
Unless $\sigma$ vanishes, there are always eigenvalues with $\mathrm{Re}(\lambda_k) > 0$. We dismiss such theories with diverging modes and impose $\sigma = 0$. It turns out (see Appendix \ref{linear-eom}) that every divergence in the equations of motion is due to the same combination of gravitational constants. Setting this combination to zero reduces the number of gravitational constants to 7 and defines the sector of linear area metric gravity theories with physically relevant phenomenology.

\subsection{Relation to canonical gravitational closure}
Remarkably, this linear theory is equivalent to the linear theory obtained by means of canonical gravitational closure \cite{Giesel_2012,Schuller_2014,D_ll_2018,Schneider_2017,Alex_2019}. However, there are differences between both approaches worth being highlighted: While it has been claimed \cite{Giesel_2012,Schuller_2014,D_ll_2018} that canonical gravitational closure rests on the principal of reconciling gravity causality with matter causality, we argue that causal compatibility is \emph{not} inherent in this approach. The mere fact that canonical gravitational closure solves the gravitational constraint algebra using a specific frame defined by matter causality does not restrict the gravitational theory to this causality (see also \cite{Reinhart_2019}). The constraint algebra is a manifestation of diffeomorphism invariance, the solution to which---consequently---yields a diffeomorphism invariant theory. It is due to the coincidence pointed out by Theorem \ref{poly-match} that both causalities coincide to first order if diffeomorphism invariance is implemented on the gravity side.

However, the linear theory obtained by canonical closure in \cite{Schneider_2017} does not even exhibit the same causality as GLED \emph{unless a gravitational constant is fixed} \cite{Alex_2019}. This hints at missing equations constraining the linear theory properly and we suspect these equations to be equivalent to those implementing Lorentz invariance of the perturbation coefficients in the covariant approach \cite{Alex_2020}. As these equations are obtained only after a prolongation of the system, it is natural for them to be missing in \cite{Schneider_2017}. Using Lorentz invariant ansätze circumvents the need for additional equations, a fact that has been exploited for the present work. In \cite{Schneider_2017}, spatial ansätze built from $\gamma$ and $\epsilon$ have been constructed for the 3+1 theory, effectively implementing an $O(3)$ symmetry. The discrepancy between $O(1,3)$ invariant spacetime ansätze for the covariant theory and $O(3)$ invariant spatial ansätze for the 3+1 theory is fixed by said choice of a gravitational constant. This is symptomatic for the intricacies that come with the infinity of canonical closure equations as compared to the 137 equivariance equations: A PDE theoretic analysis, which is necessary in order to devise a perturbative solution strategy, is much more complicated in the former case, while in the latter case, it comes almost for free.

Linear area metric gravity as constructed in the canonical picture \cite{Schneider_2017} is the basis for predictions in e.g.~lensing \cite{Schuller_2017}, quantum electrodynamics \cite{GrosseHolz_2017}, or galactic dynamics \cite{Rieser_2020}. Our findings support these predictions, as they make use of a complementary approach and still provide the same theoretical basis, while also addressing some question marks as pointed out above.

\section{Gravitational radiation from a binary system}
\label{binary_system}
A dynamical theory for matter which makes use of some geometry is always incomplete as long as the dynamics of the geometry are not known. Gravity closes this picture by providing the missing link. Only the joint model of matter theory and gravity enables the physicist to predict the evolution of matter over time---while also predicting how geometry evolves in the process.

In this final part of the present work, we make use of second-order area metric gravity as derived above in order to demonstrate how covariant constructive gravity completes a matter theory to a joint theory of matter and gravity by predicting a nontrivial interaction: the generation of gravitational waves from a gravitationally bound matter distribution.

\subsection{Iterative solution strategy}

Let the matter in question be a field $\phi$ in some bundle over spacetime and the geometry be a field $G$ in some other spacetime bundle. $G$ provides the local structure necessary to formulate the matter action $S_\text{matter}\lbrack\phi,G)$. Covariant constructive gravity yields the total action
\begin{equation}\label{total-action}
  S\lbrack G,\phi\rbrack = S_\text{gravity}\lbrack G\rbrack + \kappa S_\text{matter}\lbrack\phi,G)
\end{equation}
by providing the gravity action $S_\text{gravity}\lbrack G\rbrack$. The coupling constant $\kappa$ controls the scale of coupling between matter and geometry. Variations w.r.t.~both fields yield the Euler-Lagrange equations
\begin{equation}\label{total-eom}
  \begin{aligned}
    e\lbrack G\rbrack = {} & - \kappa T\lbrack\phi,G), \\
    f\lbrack \phi,G) = {} & 0,
  \end{aligned}
\end{equation}
with abbreviations
\begin{equation}\label{total-eom-parts}
  e\lbrack G\rbrack=\frac{\delta S_\text{gravity}}{\delta G},\quad T\lbrack\phi,G)=\frac{\delta S_\text{matter}}{\delta G},\quad f\lbrack\phi,G)=\frac{\delta S_\text{matter}}{\delta \phi}
\end{equation}
for the constituents.

The PDE system (\ref{total-eom}) is, in general, tightly coupled and correspondingly hard to solve. Effects of finite order in the coupling can, however, be calculated by perturbative iteration. We proceed similarly as in \cite{poisson2014gravity} and expand the geometry formally as
\begin{equation}\label{formal-expansion}
  G = N + \sum_{k=1}^{\infty} \kappa^k H_{(k)}.
\end{equation}
Truncating (\ref{formal-expansion}) at order $k$ yields an approximation $G_{(k)}$ of the geometric field $G$. We also expand the contributions $e$ and $T$ to the Euler-Lagrange equations,
\begin{equation}
  \begin{aligned}
    e\lbrack N+H\rbrack = {} & e_{(0)} + e_{(1)}\lbrack H\rbrack + e_{(2)}\lbrack H\rbrack + \mathcal O(H^3), \\
    T\lbrack\phi,N+H) = {} & T_{(0)}\lbrack\phi\rbrack + T_{(1)}\lbrack\phi,H) + \mathcal O(H^2).
  \end{aligned}
\end{equation}

For the \emph{zeroth} iteration, we evaluate (\ref{total-eom}) at $G_{(0)}=N$, which yields the equation $e\lbrack N\rbrack = e_{(0)} = 0$ for the zeroth order. This states that $N$ has to be chosen as solution to the gravitational field equations \emph{in vacuo}.

The \emph{first} iteration starts with evaluating the Euler-Lagrange equations at $G_{(1)} = N + \kappa H_{(1)}$. Using $e_{(0)} = 0$ from the zeroth iteration, this yields an equation for $H_{(1)}$
\begin{equation}
  e_{(1)}\lbrack H_{(1)}\rbrack = - T_{(0)}\lbrack\phi\rbrack.
\end{equation}
From the solution $H_{(1)}$, we fix the matter field to first order by solving $f\lbrack\phi,G_{(1)}) = 0 + \mathcal O(\kappa^2)$ for $\phi$.

The \emph{second} iteration builds up on this result. We insert the expansion $G_{(2)} = N + \kappa H_{(1)} + \kappa^2 H_{(2)}$ in (\ref{total-eom}), make use of the lower-order equations for $N$ and $H_{(1)}$ and obtain for $H_{(2)}$
\begin{equation}\label{second-order}
  e_{(1)}\lbrack H_{(2)}\rbrack = -\kappa^{-1} T_{(0)}\lbrack\phi\rbrack - T_{(1)}\lbrack\phi,H_{(1)}) - e_{(2)}\lbrack H_{(1)}\rbrack + \mathcal O(\kappa).
\end{equation}
Note that $\phi$ depends on $\kappa H_{(1)}$, so we have to be careful to only consider terms of order $\kappa^1$ from $T_{(0)}\lbrack\phi\rbrack$ and of order $\kappa^0$ from $T_{(1)}\lbrack\phi,H_{(1)})$ when solving (\ref{second-order}).

Aborting the procedure at this point, the final result is an approximation $G_{(2)}=N+\kappa H_{(1)}+\kappa^2 H_{(2)}$ of the geometry sourced by matter $\phi$ subject to linearized gravity.

\subsection{Einstein gravity}
\label{metric_binary_system}
Let us first apply the iterative solution strategy to a binary system subject to Einstein gravity.\footnote{The result is of course well known and extends to much more complex configurations of matter as well as higher orders in the perturbation, see \cite{poisson2014gravity} for a modern treatment. This section is concerned with developing an approach hand-tailored to the binary system and easy to adapt to area metric gravity.} The matter content of spacetime is given by two slowly moving point masses $m_i$ following two world lines $\gamma_{(i)}\colon \mathbb R \rightarrow M$. The spacetime metric field $g\in\Gamma(T^0_2M)$ measures the length of the world lines and thus provides the action
\begin{equation}\label{metric-pp-action}
  S_\text{matter}\lbrack\gamma_{(1)},\gamma_{(2)},g) = \sum_{i=1,2} m_i c \int \mathrm d\lambda \sqrt{g(\dot\gamma_{(i)}(\lambda), \dot\gamma_{(i)}(\lambda))}.
\end{equation}
The dynamical theory for the geometry $g$ completing (\ref{metric-pp-action}) to a model with predictive power is Einstein's general relativity with the Einstein-Hilbert action
\begin{equation}\label{eh-action}
  S_\text{gravity}\lbrack g\rbrack = \frac{c^3}{16\pi G} \int \mathrm d^4x \sqrt{-g} R.
\end{equation}
Performing the variations (\ref{total-eom-parts}) and using the parameterization $\gamma_{(i)}^0(\lambda) = ct$, we get the Euler-Lagrange equations
\begin{equation}\label{einstein-eqn}
  \sqrt{-g}\big\lbrack R^{ab} - \frac{1}{2} g^{ab} R\big\rbrack = \frac{8\pi G}{c^3} \sum_{i=1,2} m_i \delta^{(3)}(\vec x - \vec\gamma_{(i)}(t)) \frac{\dot\gamma_{(i)}^a \dot\gamma_{(i)}^b}{\sqrt{g(\dot\gamma_{(i)},\dot\gamma_{(i)})}}
\end{equation}
and
\begin{equation}\label{geodesic-eqn}
  0 = \ddot\gamma_{(i)}^a + \Gamma^{a}_{\hphantom abc} \dot\gamma_{(i)}^b \dot\gamma_{(i)}^c.
\end{equation}
Incorporating the slow movement of the source as $\dot\gamma_{(i)}^\alpha/c\ll1$ simplifies (\ref{geodesic-eqn}) to
\begin{equation}\label{geodesic-eqn-slow}
  \dot\gamma_{(i)}^0 = c \quad \text{and} \quad \frac{1}{c^2} \ddot\gamma_{(i)}^\alpha = -\Gamma^{\alpha}_{\hphantom\alpha00}.
\end{equation}

We now construct the perturbative solution to second order around the Minkowski metric, i.e., $g^{ab} = \eta^{ab} + h^{ab} = \eta^{ab} + G h_{(1)}^{ab} + G^2 h_{(2)}^{ab} + \mathcal O(G^3)$. The perturbation decomposes in the usual way as (compare Eq.~\ref{metric-decomposition})
\begin{equation}
  h^{00} = -2A,\quad h^{0\alpha} = B^\alpha,\quad h^{\alpha\beta} = -E^{\alpha\beta} - \gamma^{\alpha\beta} C.
\end{equation}
$E^{\alpha\beta}$ is a transverse traceless tensor and $B^\alpha$ a transverse vector as introduced in Sec.~\ref{gauge_fixing}. We made use of a gauge condition which sets the scalar part of $h^{0\alpha}$ as well as the vector and traceless scalar part of $h^{\alpha\beta}$ to zero.

The zeroth iteration is already solved because the left-hand side of (\ref{einstein-eqn}) evaluated at $g_{(0)} = \eta$ amounts to zero.

For the first iteration, we expand the left-hand side of (\ref{einstein-eqn}) to first order, which yields the decomposition
\begin{equation}
  \begin{aligned}
    e_{(1)}^{00}\lbrack h\rbrack = {} & \Delta C, \\
    e_{(1)}^{0\alpha}\lbrack h\rbrack = {} & -\frac{1}{2} \Delta B^\alpha - \partial^\alpha \dot C, \\
    e_{(1)}^{\alpha\beta}\lbrack h\rbrack = {} & -\frac{1}{2} E^{\alpha\beta} + \partial^{(\alpha} \dot B^{\beta)} + \gamma^{\alpha\beta} \lbrack \ddot C - \frac{2}{3} \Delta (-A + \frac{1}{2} C)\rbrack + \Delta^{\alpha\beta} \lbrack -A + \frac{1}{2} C\rbrack.
  \end{aligned}
\end{equation}
Expanding also the right-hand side to first order, incorporating the slow-motion condition, and evaluating at $g_{(1)} = \eta + G h_{(1)}$ gives the equations
\begin{equation}
  \begin{aligned}
    e_{(1)}^{00}\lbrack h_{(1)}\rbrack = {} & \frac{8\pi}{c^2} \underbrace{\sum_{i=1,2} m_i \delta^{(3)}(\vec x - \vec\gamma_{(i)}(t))}_{\rho(\vec x,t)}, \\
    e_{(1)}^{0\alpha}\lbrack h_{(1)}\rbrack = {} & 0, \\
    e_{(1)}^{\alpha\beta}\lbrack h_{(1)}\rbrack = {} & 0.
  \end{aligned}
\end{equation}
Because much of this PDE system is trivial, the solution is composed of only one scalar potential $\phi$ such that
\begin{equation}\label{metric-solution-first}
  \begin{gathered}
    E_{(1)}^{\alpha\beta} = 0,\quad B_{(1)}^\alpha = 0,\quad A_{(1)}=\phi/c^2,\quad C_{(1)}=2\phi/c^2, \\
    \begin{aligned}
      \phi(\vec x,t) = {} & -\int\mathrm d^3 \vec y \frac{\rho(\vec y,t)}{\lvert \vec x-\vec y\rvert} \\
      = {} & -\frac{m_1}{\lvert\vec x-\vec\gamma_{(1)}(t)\rvert} - \frac{m_2}{\lvert\vec x-\vec\gamma_{(2)}(t)\rvert}.
    \end{aligned}
  \end{gathered}
\end{equation}
On this linear background, we now solve the geodesic equations (\ref{geodesic-eqn}). Doing so, we encounter a common problem with the point mass idealization. The gravitational field sourced by a point mass diverges at its location. Thus, divergences arise whenever a mass ``feels'' its own field. There are two remedies pointed out in \cite{poisson2014gravity}: Either give up the idealization and model the masses as extended fluids or perform a regularization of the diverging integrals like (\ref{metric-solution-first}). Both effectively boil down to the same rule of thumb: We can keep using the point mass idealization, but must discard the diverging terms. With this in mind, we obtain the equations of motion
\begin{equation}\label{newtonian-approx}
  \ddot\gamma_{(i)}^\alpha = - G \sum_{j\neq i} m_j \frac{\gamma_{(i)}^\alpha - \gamma_{(j)}^\alpha}{\lvert\vec\gamma_{(i)} - \vec\gamma_{(j)}\rvert^3}.
\end{equation}
This is of course the Newtonian limit of general relativity. We know from Newtonian mechanics that the solutions to (\ref{newtonian-approx}) are conic sections. For our purposes, we keep it simple and consider circular solutions with constant separation $r$ and trajectories
\begin{equation}
  \vec\gamma_{(1)}(t) = \frac{m_2}{m} r \vec n,\quad\vec\gamma_{(2)}(t) = -\frac{m_1}{m} r \vec n,
\end{equation}
where $m=m_1+m_2$ is the total mass of the system and the angular frequency amounts to $\omega^2 = Gm/r^3$. The vectors $\vec n$ and $\vec\lambda$ (used below) are, in an orbit-adapted frame\footnote{An orbit-adapted frame \cite{poisson2014gravity} consists of two perpendicular vectors spanning the orbital plane and a third vector perpendicular to this plane.}, given as
\begin{equation}\label{source-frame}
  \vec n = \begin{pmatrix} \cos \omega t \\ \sin \omega t \\ 0 \end{pmatrix},\quad\vec \lambda = \begin{pmatrix} -\sin \omega t \\ \cos \omega t \\ 0 \end{pmatrix}.
\end{equation}

We have now set the stage for the second iteration. As our goal is to predict the generation of gravitational waves, we concern ourselves with the transverse traceless modes of the perturbation---the other modes do not propagate and thus cannot radiate. To this end, we expand (\ref{einstein-eqn}) to second order and evaluate at $g_{(2)} = \eta + G h_{(1)} + G^2 h_{(2)}$, which gives the relevant equation for the transverse traceless mode\footnote{The gravitational constant in the denominator is canceled by a gravitational constant arising from the time derivatives of the world lines.}
\begin{equation}\label{metric-second-iteration}
  e_{(1)}^{\alpha\beta}\lbrack h_{(2)}\rbrack = \frac{8\pi}{c^2} \sum_{i=1,2} m_i \delta^{(3)}(\vec x - \vec\gamma_{(i)}(t)) \frac{\dot\gamma_{(i)}^\alpha\dot\gamma_{(i)}^\beta}{Gc^2} - e_{(2)}^{\alpha\beta}\lbrack h_{(1)}\rbrack.
\end{equation}
The fact that the second order $e_{(2)}$ of $e$ is evaluated at the result (\ref{metric-solution-first}) of the first iteration comes in very useful because $e_{(2)}\lbrack h_{(1)}\rbrack$ can only contain the scalar potential $\phi$. Expanding the Einstein tensor to second order, the transverse traceless projection (denoted by $\lbrack{}\cdot{}\rbrack^\text{TT}$) turns out to be
\begin{equation}
  e_{(2)}^{\alpha\beta}\lbrack h_{(1)}\rbrack^\text{TT} = \frac{1}{c^4} \left\lbrack -2\partial^\alpha\phi\partial^\beta\phi\right\rbrack^\text{TT} + \frac{1}{c^4}\left\lbrack 4\partial^\alpha(\phi\partial^\beta\phi)\right\rbrack^\text{TT}.
\end{equation}
Using this expansion in (\ref{metric-second-iteration}), the field equation for the transverse traceless mode reads
\begin{equation}\label{metric-second-iteration-pde}
  \begin{aligned}
    \Box E_{(2)}^{\alpha\beta} = {} & -\frac{16\pi}{Gc^4}\bigg\lbrack\sum_{i=1,2}m_i\delta^{(3)}(\vec x-\vec\gamma_{(i)}(t))\dot\gamma_{(i)}^\alpha\dot\gamma_{(i)}^\beta\bigg\rbrack^\text{TT} \\
                           {} & -\frac{4}{c^4}\left\lbrack \partial^\alpha\phi\partial^\beta\phi - 2 \partial^\alpha(\phi\partial^\beta\phi)\right\rbrack^\text{TT}.
  \end{aligned}
\end{equation}
The retarded solution to a wave equation of the kind $\Box\psi(\vec x,t) = 4\pi \varphi(\vec x,t)$ is obtained by convolution of the source with the retarded Green's function,
\begin{equation}\label{green}
  \psi(\vec x,t) = \int\mathrm d^3\vec y\,\frac{\varphi(\tau,\vec y)}{\lvert\vec x-\vec y\rvert},
\end{equation}
where $\tau = t-\lvert\vec x-\vec y\rvert/c$ is the retarded time. As we are only interested in radiation into the far zone $R=\lvert\vec x\rvert\gg r$ and the sources in (\ref{metric-second-iteration-pde}) are either confined to a bound region of radius $r$---the separation of the binary system---or decreasing with $1/\lvert\vec y\rvert^4$, the zeroth order in the expansion of $\lvert\vec x-\vec y\rvert$ in (\ref{green}) is a first approximation, such that
\begin{equation}\label{wave-approx}
  \psi(\vec x,t) = \frac{1}{R} \int\mathrm d^3\vec y\,\varphi(\tau,\vec y)
\end{equation}
and $\tau=t-R/c$. This leaves us with two integrals to be evaluated, where we can already drop a boundary term in the second integral,
\begin{equation}
  \begin{aligned}
    K^{\alpha\beta} = {} & \int \mathrm d^3\vec y \, \rho(\tau,\vec y) \dot\gamma_{(i)}^\alpha \dot\gamma_{(i)}^\beta, \\
    U^{\alpha\beta} = {} & \int \mathrm d^3\vec y \, \partial^\alpha\phi\partial^\beta\phi.
  \end{aligned}
\end{equation}

For the integral $K^{\alpha\beta}$, we readily obtain
\begin{equation}\label{metric-k-sol}
  K^{\alpha\beta} = \frac{G\eta m^2}{r} \lambda^\alpha \lambda^\beta,
\end{equation}
with the reduced mass $\eta = \frac{m_1m_2}{(m_1+m_2)^2}$.

The integral $U^{\alpha\beta}$ requires more careful consideration. First, it is instructive to make use of the integral representation (\ref{metric-solution-first}) of the scalar potential $\phi$
\begin{equation}
  U^{\alpha\beta} = \int\mathrm d^3\vec y\int\mathrm d^3\vec{y^\prime}\int\mathrm d^3\vec{y^{\prime\prime}} \frac{\rho(\vec{y^\prime})\rho(\vec{y^{\prime\prime}})}{\lvert\vec y-\vec{y^\prime}\rvert^3\lvert\vec y-\vec{y^{\prime\prime}}\rvert^3} (y^\alpha - y^{\prime\alpha})(y^\beta - y^{\prime\prime\beta}).
\end{equation}
Performing the integration over $\vec y$ gives
\begin{equation}
  U^{\alpha\beta} = 2\pi \int\mathrm d^3\vec{y^\prime}\int\mathrm d^3\vec{y^{\prime\prime}}\frac{\rho(\vec{y^\prime})\rho(\vec{y^{\prime\prime}})}{\lvert\vec{y^\prime}-\vec{y^{\prime\prime}}\rvert} \left\lbrack \gamma^{\alpha\beta} - \frac{(y^{\prime\alpha}-y^{\prime\prime\alpha})(y^{\prime\beta}-y^{\prime\prime\beta})}{\lvert\vec{y^\prime}-\vec{y^{\prime\prime}}\rvert^2}\right\rbrack.
\end{equation}
This integral can now be evaluated. Leaving out diverging terms in order to regularize the integral, as already explained, results in
\begin{equation}\label{metric-u-sol}
  U^{\alpha\beta} = \frac{4\pi\eta m^2}{r} \lbrack \gamma^{\alpha\beta} - n^\alpha n^\beta\rbrack.
\end{equation}
We now put together (\ref{metric-k-sol}) and (\ref{metric-u-sol}) and remove the trace in order to obtain the far-field solution to equation (\ref{metric-second-iteration-pde}).

Finally, we can predict that, to lowest order and in the far field, a binary system of reduced mass $\eta$ and total mass $m$ with separation $r$ in circular motion emits gravitational waves as
\begin{equation}\label{metric-solution}
  G^2 h_{(2)}^{\alpha\beta} = \frac{4\eta}{c^4 R} \frac{(Gm)^2}{r} \lbrack \lambda^\alpha \lambda^\beta - n^\alpha n^\beta\rbrack^\text{TT}.
\end{equation}
This is in accordance with the results from post-Minkowskian and post-Newtonian theory in the literature \cite{poisson2014gravity}, which have been confirmed by indirect \cite{Taylor_1982,Taylor_1979,Weisberg_2016} and, recently, direct \cite{Abbott_2016,Abbott_2016_2,Abbott_2019} observations. Note that we followed a top-down approach, starting from the Einstein-Hilbert action (\ref{eh-action}) and solving perturbatively up to second order. We would have arrived at the same result using the bottom-up approach provided by perturbative covariant gravity, because the perturbative construction of metric gravity to second order in the field equations coincides with the corresponding expansion of the Einstein equations \cite{Alex_2020}. Above procedure allows for the prediction of a nontrivial second-order effect of matter-gravity interaction from a theory constructed to second order---a technique we will subsequently apply to area metric gravity, whose exact dynamics are \emph{not} known.

It is clear from our calculation that the generation of gravitational waves is indeed a second-order effect. The system is bound by gravity as a first-order effect and gravitational radiation is sourced by this gravitationally bound system as effect of second order. Derivations in the older literature that arrive at (\ref{metric-solution}) or its generalization called \emph{quadrupole formula} from linearized gravity are incorrect in their premises: Either they silently make use of the next order at some point or inadvertently construct this order along the way. Earlier results show that there is no leeway in the (iterative) construction of metric gravity \cite{Lovelock_1969,Lovelock_1971,Lovelock_1972,Hojman_1976,Deser_1970,Alex_2020}, so it comes as little surprise that, eventually, the correct formula is obtained nevertheless.

\subsection{Area metric gravity}
\label{area_metric_binary_system}
As already pointed out in Sec.~\ref{field-equations}, the action for two point masses following GLED dynamics is given as
\begin{equation}\label{area-matter-lagrangian}
  S_\text{matter}\lbrack\gamma_{(1)},\gamma_{(2)},G) = \sum_{i=1,2} m_i c \int \mathrm d\lambda P_\text{GLED}(L^{-1}(\dot\gamma_{(i)}(\lambda)))^{-\frac{1}{4}}.
\end{equation}
Up to first order in the expansion around $N$ as defined in (\ref{expansion-point}), $P_\text{GLED}$ is equivalent to a quadratic polynomial (\ref{fresnel_expansion}), which using the perturbation fields (\ref{quantities2}) amounts to
\begin{equation}
  P^{(1)}(k) = \eta(k,k) + \lbrack-2A\rbrack k_0 k_0 + \lbrack -2b^\alpha\rbrack k_0 k_\alpha + \lbrack -\frac{1}{2} u^{\alpha\beta} -\frac{1}{2} \gamma_{\mu\nu} w^{\mu\nu} \gamma^{\alpha\beta}\rbrack k_\alpha k_\beta.
\end{equation}
The causality defined by $P^{(1)}$ is effectively \emph{metric}, such that (\ref{area-matter-lagrangian}) is obtained by simple inversion (see \cite{R_tzel_2011}) as
\begin{equation}\label{area-matter-lagrangian-linear}
  \begin{aligned}
  S_\text{matter}\lbrack\gamma_{(1)},\gamma_{(2)},G\rbrack = \sum_{i=1,2} m_i c \int \mathrm d\lambda {} & \bigg\{ \eta_{ab}\dot\gamma_{(i)}^a\dot\gamma_{(i)}^b \\
    {} & + 2A \dot\gamma_{(i)}^0\dot\gamma_{(i)}^0 -2b_\alpha \dot\gamma_{(i)}^0\dot\gamma_{(i)}^\alpha \\
    {} & + \left\lbrack \frac{1}{2} u_{\alpha\beta} + \frac{1}{2} \gamma^{\mu\nu} v_{\mu\nu} \gamma_{\alpha\beta} \right\rbrack \dot\gamma_{(i)}^\alpha \dot\gamma_{(i)}^\beta\bigg\}^{\frac{1}{2}} \\
    + \mathcal O(H^2). \hspace{2em} &
  \end{aligned}
\end{equation}
In Sec.~\ref{construction}, we have constructed the gravitational Lagrangian $\mathcal L$, with the help of which we formulate the gravitational action
\begin{equation}\label{area-action}
  S_\text{gravity}\lbrack G\rbrack = \frac{c^3}{16\pi G} \int \mathrm d^4x\,\mathcal L.
\end{equation}

We now start the iterative solution procedure for a perturbation around $N$. The zeroth iteration is already solved, because the construction procedure of the gravitational dynamics has been set up such that $N$ is a vacuum solution.

For the first iteration, the perturbation $H_{(1)}$ is---due to the slow-motion condition---sourced by only one contribution,
\begin{equation}\label{area-first-order}
  \frac{\delta S_\text{matter}}{\delta A} = -\frac{16\pi G}{c^2} \rho(\vec x,t).
\end{equation}
Using again the slow-motion condition and working within the gauge (\ref{area-gauge}), the solution to the linearized area metric gravity field equations sourced by (\ref{area-first-order}) is already known from (\ref{scalar-solution}), such that the nonvanishing modes read
\begin{equation}\label{area-first-order-solution}
  \begin{aligned}
    A_{(1)} = {} & -\frac{1}{c^2} \int\mathrm d^3\vec y\rho(\vec y)\left\lbrack \frac{\alpha}{\lvert \vec x-\vec y\rvert} + \frac{\beta\mathrm e^{-\mu\lvert\vec x-\vec y\rvert}}{\lvert\vec x-\vec y\rvert}\right\rbrack \\
    \tilde V_{(1)} = {} & -\frac{1}{c^2} \int\mathrm d^3\vec y\rho(\vec y)\left\lbrack\frac{\gamma\mathrm e^{-\mu\lvert\vec x-\vec y\rvert}}{\lvert\vec x-\vec y\rvert}\right\rbrack \\
    \tilde U_{(1)} = {} & 4 A_{(1)} - (3 + 8 \frac{\beta}{\gamma}) \tilde V_{(1)}.
  \end{aligned}
\end{equation}
Evaluating the integrals, we readily get
\begin{equation}
  \begin{aligned}
    A_{(1)} = {} & -\frac{1}{c^2} \sum_{i=1,2}m_i\left\lbrack \frac{\alpha}{\lvert \vec x-\vec \gamma_{(i)}(t)\rvert} + \frac{\beta\mathrm e^{-\mu\lvert\vec x-\vec \gamma_{(i)}(t)\rvert}}{\lvert\vec x-\vec \gamma_{(i)}(t)\rvert}\right\rbrack \\
    \tilde V_{(1)} = {} & -\frac{1}{c^2} \sum_{i=1,2}m_i\left\lbrack \frac{\gamma\mathrm e^{-\mu\lvert\vec x-\vec \gamma_{(i)}(t)\rvert}}{\lvert\vec x-\vec \gamma_{(i)}(t)\rvert}\right\rbrack.
  \end{aligned}
\end{equation}
Because the matter action (\ref{area-matter-lagrangian-linear}) is given by an effective metric, the equations of motion for the source are again geodesic equations (\ref{geodesic-eqn-slow}). The relevant Christoffel symbol expands as $\Gamma^\alpha_{\hphantom{\alpha}00} = G \partial^\alpha A + \mathcal O(G^2)$. We again find the circular solutions of constant separation $r$
\begin{equation}
  \vec\gamma_{(1)}(t) = \frac{m_2}{m} r \vec n,\quad\vec\gamma_{(2)}(t) = -\frac{m_1}{m} r \vec n.
\end{equation}
This time, the angular velocity amounts to
\begin{equation}
  \omega^2 = \frac{(G\alpha) m}{r^3} \left\lbrack 1 + \frac{\beta}{\alpha} \mathrm e^{-\mu r}(1 + \mu r)\right\rbrack.
\end{equation}

Having solved the first iteration for $H_{(1)}$ and the source trajectories, we are ready to proceed with the second iteration. Let us begin with the transverse traceless modes of $H_{(2)}$, one of which is subject to the massless wave equation (\ref{tt-eqns-reduced})
\begin{equation}\label{second-order-area-tt-u}
  \begin{aligned}
    \frac{c^3G}{16\pi} \frac{1}{8\alpha} \Box U_{(2)}^{\alpha\beta} = -\bigg\{ & \left(\frac{\delta S_\text{gravity}}{\delta u_{\alpha\beta}}\right)_{(2)}\lbrack GH_{(1)}\rbrack + \frac{1}{4c} \sum_{i=1,2}m_i\delta^{(3)}(\vec x-\vec\gamma_{(i)}(t)) \dot\gamma_{(i)}^{\alpha} \dot\gamma_{(i)}^{\beta} \bigg\}^\text{TT}.
  \end{aligned}
\end{equation}
Using the third-order Lagrangian (see Appendix \ref{ansaetze} and \ref{reduction}), we perform a 3+1 split with \textsc{Cadabra}\footnote{The code is publicly available at \cite{Alex_2020_area-metric-gravity}} and evaluate the second-order gravitational field equations $(\delta S_\text{gravity}/\delta u_{\alpha\beta})_{(2)}$ at the first-order solution (\ref{area-first-order-solution}), obtaining
\begin{equation}\label{second-order-area-tt-u-1}
  \left(\frac{\delta S_\text{gravity}}{\delta u_{\alpha\beta}}\right)_{(2)}^\text{TT}\lbrack GH_{(1)}\rbrack = \frac{G}{16\pi c}\left\lbrack \alpha \partial^\alpha X\partial^\beta X + \beta \partial^\alpha Y\partial^\beta Y\right\rbrack^\text{TT},
\end{equation}
where we introduced abbreviations for the integrals
\begin{equation}
  \begin{aligned}
    X = {} & \int\mathrm d^3\vec y\rho(\vec y) \frac{1}{\lvert\vec x-\vec y\rvert}, \\
    Y = {} & \int\mathrm d^3\vec y\rho(\vec y) \frac{\mathrm e^{-\mu\lvert\vec x-\vec y\rvert}}{\lvert\vec x-\vec y\rvert}.
  \end{aligned}
\end{equation}
With (\ref{second-order-area-tt-u-1}) in (\ref{second-order-area-tt-u}), we can now integrate the wave equation using the same approximation as before (see Eq.~\ref{wave-approx}), such that the solution reads
\begin{equation}
  U_{(2)}^{\alpha\beta} = -\left\lbrack \frac{8\alpha}{Gc^4R} K^{\alpha\beta} + \frac{2\alpha^2}{\pi c^4R} \left(\Phi^{\alpha\beta} + \frac{\beta}{\alpha}\Psi^{\alpha\beta}\right) \right\rbrack^\text{TT}
\end{equation}
with the kinetic term
\begin{equation}
  K^{\alpha\beta} = \int\mathrm d^3\vec y\rho(\vec y)\dot\gamma_{(i)}^\alpha\dot\gamma_{(i)}^\beta
\end{equation}
and the potential terms
\begin{equation}
  \Phi_{\alpha\beta} = \int\mathrm d^3\vec y\int\mathrm d^3\vec{y^\prime}\int \mathrm d^3 \vec{y^{\prime\prime}} \rho(\vec{y^\prime}) \rho(\vec{y^{\prime\prime}}) \left( \partial_\alpha \frac{1}{\lvert\vec z\rvert}\right)_{\vec z=\lvert\vec y-\vec{y^\prime}\rvert} \left( \partial_\beta \frac{1}{\lvert\vec z\rvert}\right)_{\vec z=\lvert\vec y-\vec{y^{\prime\prime}}\rvert}
\end{equation}
and
\begin{equation}
  \Psi_{\alpha\beta} = \int\mathrm d^3\vec y\int\mathrm d^3\vec{y^\prime}\int \mathrm d^3 \vec{y^{\prime\prime}} \rho(\vec{y^\prime}) \rho(\vec{y^{\prime\prime}}) \left( \partial_\alpha \frac{\mathrm e^{-\mu\lvert\vec z\rvert}}{\lvert\vec z\rvert}\right)_{\vec z=\lvert\vec y-\vec{y^\prime}\rvert} \left( \partial_\beta \frac{\mathrm e^{-\mu\lvert\vec z\rvert}}{\lvert\vec z\rvert}\right)_{\vec z=\lvert\vec y-\vec{y^{\prime\prime}}\rvert}.
\end{equation}
Evaluating all three integrals, we finally obtain the gravitational radiation on the massless $U^{\alpha\beta}$ mode in the far zone from a binary system with constant separation $r$,
\begin{equation}\label{area-metric-solution}
  G^2U_{(2)}^{\alpha\beta} = -\frac{8\eta}{c^4R}\frac{(G\alpha m)^2}{r}\lbrack 1+f(r)\rbrack\lbrack \lambda^\alpha \lambda^\beta - n^\alpha n^\beta\rbrack^\text{TT},
\end{equation}
where $f(r) = \frac{\beta}{\alpha}(1+\mu r)\mathrm e^{-\mu r}$.

This result is of remarkable consequence, because with (\ref{induced_area_perturbative}) in mind, we see that for $\alpha=1$ it is the transverse traceless wave induced by the metric gravitational wave $E_{(2)}^{\alpha\beta}$ (see Eq.~\ref{metric-solution}) plus a Yukawa correction $f(r)$ that falls off exponentially with the separation,
\begin{equation}\label{area-metric-solution-correction}
  G^2U_{(2)}^{\alpha\beta} = 2G^2E_{(2)}^{\alpha\beta} \lbrack 1+f(r)\rbrack.
\end{equation}
That is, the result (\ref{area-metric-solution}) is a \emph{refinement} of the metric result, offering the same qualitative behavior with quantitative corrections given by short-ranging Yukawa terms. In particular, the phenomenology of gravitational radiation in Einstein gravity is contained within this result, either for sufficiently large separation $r$ of bodies or for appropriate choices of area metric gravitational constants.

As the inspection of the linearized theory in Sec.~\ref{field-equations} revealed, the metrically inducible mode $U^{\alpha\beta}$ is the only massless degree of freedom. All other propagating modes are subject to massive wave equations. First, let us consider the trace-free fields $\tilde v^{\alpha\beta} = v^{\alpha\beta} - 1/3\gamma^{\alpha\beta}\gamma_{\mu\nu} v^{\mu\nu}$ and $w^{\alpha\beta}$. Since the first-order matter action (\ref{area-matter-lagrangian-linear}) does only depend on the trace of $v^{\alpha\beta}$ and does not depend on $w^{\alpha\beta}$ at all, the respective equations of motion are not sourced by kinetic terms, leaving us with\footnote{$\lbrack t^{\alpha\beta}\rbrack^\text{TF} = t^{(\alpha\beta)} - \frac{1}{3}\gamma_{\mu\nu} t^{\mu\nu}\gamma^{\alpha\beta}$ is the idempotent projection on the symmetric trace-free part.}
\begin{equation}\label{massive-wave-equations}
  \begin{aligned}
    \Box \tilde v^{\alpha\beta}_{(2)} + \nu^2 \tilde v^{\alpha\beta}_{(2)} = {} & \delta \lbrack \partial^\alpha X \partial^\beta Y\rbrack^\text{TF}, \\
    \Box w^{\alpha\beta}_{(2)} + \nu^2 w^{\alpha\beta}_{(2)} = {} & \epsilon \lbrack \partial^\alpha X \partial^\beta Y \rbrack^\text{TF}.
  \end{aligned}
\end{equation}
$\delta$ and $\epsilon$ are gravitational constants arising during the construction of the third-order area metric gravity Lagrangian, the mass $\nu$ is a gravitational constant of the second-order Lagrangian (compare Appendix \ref{linear-eom}). These wave equations govern the radiation of traceless tensor modes $V^{\alpha\beta}$ and $W^{\alpha\beta}$, the vector modes $U^{\alpha}$ and $W^{\alpha}$ (via the gauge condition $U^\alpha = V^\alpha$), and the traceless scalar modes $V$ and $W$. The remaining trace-free mode---the vector $B^\alpha$---does not propagate on its own but follows from a constraint (see Appendix \ref{linear-eom}).

The retarded solution to a massive wave equation of the kind $(\Box + m^2)\psi(x) = 4\pi \varphi(x)$ is obtained by convolution of $\varphi$ with the retarded Green's function
\begin{equation}
  G_\text{ret}(x,y) = \theta(x^0-y^0) \int \frac{\mathrm d^3\vec k}{(2\pi)^3} \frac{\sin \omega_k(x^0-y^0)}{\omega_k} \mathrm e^{\mathrm i \vec k\cdot(\vec x-\vec y)},
\end{equation}
where $\omega_k = \sqrt{\lvert\vec k\rvert^2 + m^2}$. Carrying out the integrals for the wave equations (\ref{massive-wave-equations}), we arrive at two qualitatively different solutions, depending on the value of $\omega_0 \vcentcolon= 2\omega$.

For $\omega_0 < c\nu$, the gravitational fields are decaying exponentially with $R$. In the orbit-adapted frame, we have
\begin{equation}
  G^2\tilde v_{(2)}^{\alpha\beta} = \frac{\delta\eta}{c^4R} \frac{(Gm)^2}{r} g(r) \Bigg\lbrack 3 \mathrm e^{-\sqrt{\nu^2 - (\omega_0/c)^2}R} \begin{pmatrix} \cos \omega_0t & \sin \omega_0t & 0 \\ \sin \omega_0t & -\cos \omega_0t & 0 \\ 0 & 0 & 0 \end{pmatrix} + \mathrm e^{-\nu R}\begin{pmatrix} \frac{1}{2} & & \\ & \frac{1}{2} & \\ & & -1 \end{pmatrix} \Bigg\rbrack^{\alpha\beta},
\end{equation}
where
\begin{equation}
  g(r) = \frac{1 - \lbrack 1 + \mu r + \frac{1}{3}(\mu r)^2 \mathrm e^{-\mu r}\rbrack}{(\mu r)^2}.
\end{equation}
$w_{(2)}^{\alpha\beta}$ has the same solution, only with $\epsilon$ instead of $\delta$. This behavior is in line with results for the generation of gravitational waves on massive modes from nongravitationally bound systems \cite{M_ller_2018}: Below a certain threshold for the angular frequency (or related measures like energy) of the generating system, no radiation is emitted into the far zone. Also, the oscillating (but exponentially damped) part of the solution does not follow the retarded time, but the coordinate time---which is not the behavior of a wave-like solution.

In the case of $\omega_0 > c\nu$, the nonoscillating part of the solution remains unchanged. The oscillating part, however, is now radiating as
\begin{equation}\label{massive-radiating}
    G^2\tilde v_{(2)}^{\alpha\beta} = \frac{3 \delta \eta}{c^4R} \frac{(Gm)^2}{r} g(r) \begin{pmatrix} \cos\varphi & \sin\varphi & 0 \\ \sin\varphi & -\cos\varphi & 0 \\ 0 & 0 & 0 \end{pmatrix}^{\alpha\beta},
\end{equation}
where
\begin{equation}
      \varphi =\omega_0 t - \tilde\omega\frac{R}{c} \quad \text{and} \quad \tilde\omega = \sqrt{\omega_0^2 - (c\nu)^2}.
\end{equation}
Again, the same holds for $w_{(2)}^{\alpha\beta}$ with $\epsilon$ instead of $\delta$.

It remains to derive wave equations for the trace modes $\tilde U$, $\tilde V$, and $A$. One such equation is obtained as (see Appendix \ref{linear-eom}, Eq.~\ref{scalar-eqns-reduced})
\begin{equation}\label{massive-trace-equation}
  \Box \tilde V_{(2)} + \mu^2 \tilde V_{(2)} = -\gamma \left\lbrack \frac{1}{4} \rho_A - (1 + \frac{3}{4}\frac{\gamma}{\beta})\rho_u + \frac{\gamma}{4\beta}\rho_v\right\rbrack.
\end{equation}
$\rho_A$ denotes the variations with respect to $A$
\begin{equation}
  \begin{aligned}
    \rho_A = {} & \left( \frac{\delta S_\text{gravity}}{\delta A} \right)_{(2)}\lbrack H_{(1)}\rbrack \\
    & + \frac{16\pi}{Gc^3} \left( \frac{\delta S_\text{matter}}{\delta A}\right)_{(0)}\lbrack\gamma_{(1)},\gamma_{(2)},N\rbrack \\
    & + \frac{16\pi}{c^3} \left( \frac{\delta S_\text{matter}}{\delta A}\right)_{(1)}\lbrack\gamma_{(1)},\gamma_{(2)},H_{(1)}\rbrack \\
    & + \mathcal O(G),
  \end{aligned}
\end{equation}
$\rho_u$ and $\rho_v$ denote traces of variations w.r.t.~$u_{\alpha\beta}$ and $v_{\alpha\beta}$, respectively. In setting up the wave equations for $U_{(2)}^{\alpha\beta}$, $\tilde v_{(2)}^{\alpha\beta}$, and $w_{(2)}^{\alpha\beta}$, we did not pick up contributions from $(\delta S_\text{matter}/\delta G)_{(1)}$, because the time derivatives of the spatial trajectories, $\dot\gamma_{(i)}^\alpha$, are already of order $\mathcal O(G)$. However, an expansion of the GLED principal polynomial to second order reveals that the variation w.r.t.~$A$ comes with $\lbrack\dot\gamma_{(i)}^0\rbrack^4 = c^4 = \mathcal O(G^0)$ and thus has to be considered as source for the second iteration.

Unfortunately, the expansion of $S_\text{matter}$ to second order is not as readily accessible as the linearized action. This is because only the first-order GLED principal polynomial factors into the square of a polynomial of second degree as in (\ref{fresnel_expansion}). We leave the calculation of $\rho_A$ and the solution for $\tilde V_{(2)}$ open for future research. However, $\tilde V$ is the only propagating trace degree of freedom, because the system must have four constraint equations\footnote{This follows from the same arguments that can be made for general relativity \cite{Reinhart_2019,Alex_2020_PhD}.}, two of which already constrain $B^{\alpha}$ and the other two, consequently, must constrain the remaining degrees of freedom, $A$ and $\tilde U$.

\subsection{A simple detector}
The effect of the previously derived gravitational waves on test matter is best demonstrated using a spherical distribution of freely falling point masses as detector and considering its deformation as the wave passes through. We call this arrangement a \emph{geodesic sphere}. As the dynamics of point masses are, to first nontrivial order, given by an effective metric (\ref{area-matter-lagrangian-linear}), the standard procedure of metric geodesic deviation can be used to derive
\begin{equation}\label{deviation-ricci}
  \frac{1}{c^2}\ddot X^{\alpha} = - R^{\alpha}_{\hphantom\alpha0\beta0} X^{\beta},
\end{equation}
where $R$ is the Ricci tensor related to the effective metric and $\vec X$ is the spatial deviation vector between two test masses. Using the $3+1$ split (\ref{metric-decomposition}) of the effective metric, the deviation equation (\ref{deviation-ricci}) becomes
\begin{equation}
  \ddot X^{\alpha} = -\frac{1}{2} \left\lbrack \ddot\varphi^{\alpha}_{\hphantom\alpha\beta} + c (\dot b^\alpha_{\hphantom\alpha,\beta} + \dot b_{\beta,}^{\hphantom{\beta,}\alpha}) + 2 c^2 A_{,\hphantom\alpha\beta}^{\hphantom,\alpha}\right\rbrack X^\beta.
\end{equation}
A purely spatial perturbation has only contributions from $\varphi^{\alpha\beta}$, in which case the deviation equation is---for small deviations---easily integrated as
\begin{equation}\label{deviation-solution}
  X^{\alpha}(t) = X^{\alpha}(0) - \frac{1}{2} \varphi^{\alpha}_{\hphantom\alpha\beta}(t) X^{\beta}(0).
\end{equation}
Let us now consider the individual modes of gravitational radiation and their effects on the geodesic deviation (\ref{deviation-solution}) of test matter. All modes\footnote{Except for the trace modes, which we left open for future consideration.} are proportional to the projections of
\begin{equation}\label{generic-solution-orbit-adapted}
  M^{\alpha\beta} = \begin{pmatrix} \cos(\varphi) & \sin(\varphi) & 0 \\ \sin(\varphi) & -\cos(\varphi) & 0 \\ 0 & 0 & 1 \end{pmatrix}^{\alpha\beta}
\end{equation}
onto the respective (transverse traceless, vector, scalar traceless) subspaces. The phase $\varphi$ is defined as in (\ref{massive-radiating}) and simplifies to $\varphi = 2\omega\tau$ for massless modes.
\begin{figure}
  \includegraphics[width=.6\linewidth]{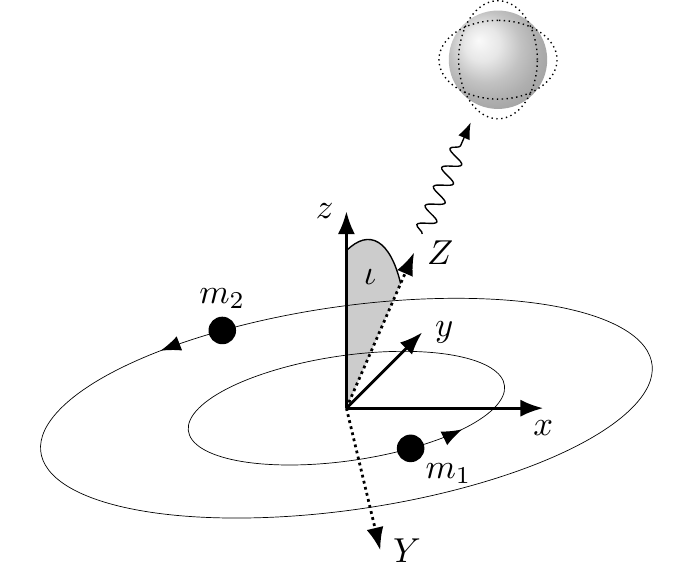}
  \caption{Orbit-adapted frame $(x,y,z)$ and detector-adapted frame $(X=x,Y,Z)$. The constituents $m_1$ and $m_2$ of the binary system describe circular orbits in the $x$-$y$ plane, producing gravitational radiation. The detector---a spherical distribution of test masses in $Z$ direction---is undergoing periodic deformations as gravitational radiation passes through. $\iota$ measures the inclination of the orbital plane with respect to the $X$-$Y$ plane.}\label{figure_1}
\end{figure}
Note that (\ref{generic-solution-orbit-adapted}) is still expressed in the \emph{orbit-adapted} frame. We now switch into a \emph{detector-adapted} frame \cite{poisson2014gravity} as illustrated in Fig.~\ref{figure_1}: The origin lies at the barycenter of the binary system, the $Z$ direction is pointing towards the spherical point mass distribution and the $X$-$Y$ plane is perpendicular to this direction. Because we only consider circular binary systems without any distinguished points on the orbits, we are free to choose the orbit-adapted $y$ direction such that the test masses lie in the $y$-$z$ plane. The detector-adapted frame is given by a simple rotation around the $x$ axis,
\begin{equation}
  \vec e_X = \begin{pmatrix} 1 \\ 0\\ 0 \end{pmatrix},\quad \vec e_Y = \begin{pmatrix} 0 \\ \cos\iota \\ -\sin\iota\end{pmatrix},\quad \vec e_Z = \begin{pmatrix} 0 \\ \sin\iota \\ \cos\iota\end{pmatrix},
\end{equation}
with the \emph{inclination angle} $\iota$. Making the transition into this frame, the tensor (\ref{generic-solution-orbit-adapted}) transforms accordingly and decomposes into the traceless tensor part
\begin{equation}\label{tt-part}
  M^{\text{TT}} = \begin{pmatrix}
    \frac{1}{2}(1+\cos^2\iota)\cos\varphi & \cos\iota\sin\varphi & 0 \\
    \cos\iota\sin\varphi & -\frac{1}{2}(1+\cos^2\iota)\cos\varphi & 0 \\
    0 & 0 & 0
  \end{pmatrix},
\end{equation}
the vector part
\begin{equation}
  M^{\text{V}} = \begin{pmatrix}
    0 & 0 & \sin\iota\sin\varphi \\
    0 & 0 & -\cos\iota\sin\iota\cos\varphi \\
    \sin\iota\sin\varphi & -\cos\iota\sin\iota\cos\varphi & 0
  \end{pmatrix},
\end{equation}
and the trace-free scalar part
\begin{equation}
  M^{\text{S-TF}} = \sin^2\iota\cos\varphi \begin{pmatrix}
    \frac{1}{2} & 0 & 0 \\
    0 & \frac{1}{2} & 0 \\
    0 & 0 & -1
  \end{pmatrix}.
\end{equation}
The traceless tensor part (\ref{tt-part}) is responsible for periodic deformations of the geodesic sphere into ellipsoids by contractions and expansions in both lateral directions $X$ and $Y$. This kind of deformation---and only this kind---arises in metric gravity, where the spatial effective metric perturbation $\varphi^{\alpha\beta}$ in (\ref{deviation-solution}) is given by the spatial part of the spacetime metric. The oscillating deformation $d\times M^\text{TT}$ is of amplitude
\begin{equation}\label{metric-tt-amplitude}
  d = \frac{2\eta}{c^4R} \frac{(Gm)^2}{r}.
\end{equation}
and, because of the gravitational wave being massless, follows the phase
\begin{equation}
  \varphi = 2\omega\tau = 2\omega (t - \frac{R}{c}).
\end{equation}
Predictions of this effect have been made since the early days of general relativity \cite{Einstein_1916,Einstein_1918} and recently confirmed in earth-bound experiments \cite{Abbott_2016,Abbott_2016_2,Abbott_2019}.

In area metric gravity, $\varphi^{\alpha\beta}$ can be read off from the point mass action (\ref{area-matter-lagrangian-linear}) as
\begin{equation}\label{effective-metric}
  \varphi^{\alpha\beta} = -\frac{1}{2}\lbrack u^{\alpha\beta} + \gamma^{\alpha\beta}\gamma_{\mu\nu} v^{\mu\nu}\rbrack.
\end{equation}
On the transverse traceless mode, the deformation of geodesic spheres corresponding to (\ref{effective-metric}) coincides with the result from metric gravity, up to a correction factor of $\lbrack1 + f(r)\rbrack$ as introduced in (\ref{area-metric-solution-correction}). This is a quantitative refinement of the metric result (\ref{metric-tt-amplitude}), which does not introduce new qualitative behavior and can be arbitrarily close to the metric result for appropriate choices of separation $r$ or gravitational constants.

The solution of the wave equation (\ref{massive-wave-equations}) for $\tilde v_{(2)}^{\alpha\beta}$, together with the gauge conditions relating $U$ and $U^\alpha$ with $-V$ and $V^\alpha$, yields the combined vector and scalar traceless contribution. For $2\omega < c\nu$, the binary system does not radiate on these modes and test matter remains unaffected. If $2\omega > c\nu$, however, radiation is switched on and deforms geodesic spheres according to the deformation matrix
\begin{equation}\label{vector-scalar-deformation}
  e \times \left\lbrack M^\text{V} - M^\text{S-TF}\right\rbrack
\end{equation}
with amplitude
\begin{equation}
  e = \frac{3\delta\eta}{4c^4R} \frac{(Gm)^2}{r}g(r) = \frac{3}{8}\delta g(r) d
\end{equation}
and phase
\begin{equation}
  \varphi = 2\omega t - \sqrt{(2\omega)^2 - (c\nu)^2}\frac{R}{c}.
\end{equation}
It is worth noting that (\ref{vector-scalar-deformation}) vanishes for $\iota=0$, the case of the detector being placed exactly along on the rotation axis of the binary system. In this configuration, the source can only induce lateral deformations of the test mass distribution, which is not at all surprising considering the geometry of this particular situation. Let us consider a second case of $\iota=\pi/2$, where the geodesic sphere lies within the orbital plane. The radiation on the transverse traceless mode is now restricted to the $+$ polarization (as defined in \cite{poisson2014gravity}). On the mixed vector and scalar trace-free mode, we have deformations in all three directions\footnote{All eigenvalues are nonzero.} according to the deformation matrix
\begin{equation}
  M = e \times \begin{pmatrix} -\frac{1}{2}\cos\varphi & 0 & \sin\varphi \\ 0 & -\frac{1}{2}\cos\varphi & 0 \\ \sin\varphi & 0 & \cos\varphi\end{pmatrix}.
\end{equation}
This constitutes new \emph{qualitative} behavior.

With the solution of the wave equation for the trace mode $\tilde V$ still pending, it is not possible to predict the exact behavior of deformations mediated via this mode. However, being proportional to the identity map, we can already infer that they are uniform scalings of geodesic spheres. The mass of the trace mode $\tilde V$ is, according to its wave equation (\ref{massive-trace-equation}), given by $\mu$, so we expect a similar low-energy cutoff as for the trace-free massive modes of mass $\nu$.

\section{Conclusions}
\label{conclusions}

We have implemented the perturbative covariant constructive gravity program for GLED compatible area metric gravity up to second order in the equations of motion, as already outlined in \cite{Alex_2020}. This included a proof of the fact that, up to this order, the causality of gravity is already fixed to GLED causality by the requirement of diffeomorphism invariance alone. A subsequent 3+1 split of the corresponding field equations exposed unphysical artifacts of the theory, which we ruled out by considering a sector of the theory where 3 of the 10 first-order gravitational constants are fixed.

With this gravitational theory at hand, we followed an iterative solution strategy to obtain the circular orbits of a binary system in linearized area metric gravity and its gravitational radiation as second-order effect. The result is a refinement of the gravitational waves emitted by a circular binary system in Einstein gravity: The two massless wave modes in area metric gravity correspond to the two propagating modes in Einstein gravity and their emission from the binary system follows the same formula (\ref{area-metric-solution-correction}), up to a correction factor determined by gravitational constants from the first-order field equations and the separation of the binary system. The remaining trace-free modes are massive and thus only generated above a certain energy threshold. Once this threshold is exceeded, the emitted radiation is described by a similar formula (\ref{massive-radiating}) as for the massless modes, but scaled with two gravitational constants coming from the second-order field equations. There are technical hurdles barring us from obtaining exact results for the trace modes in the same manner, but based on the wave equation (\ref{massive-trace-equation}) we conjecture that their generation is very similar to the other massive modes, where the mass $\nu$ is to be replaced with the mass $\mu$.

Lastly, we modeled a detector for gravitational waves as a sphere of freely falling point masses. Due to the massless transverse traceless modes emitted from the binary system, this sphere undergoes the same volume-preserving lateral deformations into ellipsoids as already known from Einstein gravity, only corrected with the above-mentioned factor. A qualitatively new kind of deformation (\ref{vector-scalar-deformation}) into all three spatial directions, which is still preserving volume, is caused by the radiation on the remaining trace-free modes. The deformations from the trace modes, of which we do not know the precise wave form at the time being, can be distinguished from the rest because they consist of uniform scalings.

Our work demonstrates the potential of covariant constructive gravity in modified gravity research: Modeling matter using nonmetric geometries necessitates the conception of a novel gravitational theory. Covariant constructive gravity establishes a procedure for the construction of such a theory. For applications where only weak geometric fields are relevant, the construction can be performed perturbatively and aborted at any order. The such obtained theory allows the prediction of quantitatively and qualitatively new phenomenology, which can in turn be used to constrain parameters or outright falsify the theory.

Several effects in linearized area metric gravity derived by canonical gravitational closure \cite{Giesel_2012,Schuller_2014,D_ll_2018,Schneider_2017} have already been described: The authors of \cite{Schuller_2017,GrosseHolz_2017} predict effects from the area-metric corrected linearized Schwarzschild solution on (quantum) electrodynamics. Galactic dynamics building up on this solution have also been investigated \cite{Rieser_2020}. Furthermore, the linearized theory is sufficient in order to predict the generation of gravitational waves from nongravitationally bound systems \cite{M_ller_2018}. In the present work, we propose an experiment for testing self-coupling in area metric gravity by predicting how orbiting point masses bound by gravity affect distributions of test masses at a large distance. Note, however, that this proposal should not be understood too literally: Realistic astrophysical sources responsible for the strong signals that can be measured with contemporary technology are much more complex than the simple configuration we considered in Sec.~\ref{binary_system}. Consequently, a prediction of the emitted waveform needs a more thorough treatment, e.g., modeling the masses as extended fluids and considering higher orders in the post-Newtonian expansion \cite{poisson2014gravity}. We kept the technical difficulty at a minimum and were still able to deduce a nontrivial effect of gravitational self-coupling. This demonstrates the predictive power of covariant constructive gravity, which should be explored further to give more detailed predictions for astrophysical measurements in promising modified theories of gravity such as area metric gravity.

\newpage

\appendix

\section{Ansätze}
\label{ansaetze}

Displayed below are the Lorentz invariant perturbation ansätze for the third-order area metric Lagrangian (\ref{expansion}). The source code for ansatz generation is publicly available \cite{Alex_2020_area-metric-gravity} and is based on two Haskell libraries \cite{Reinhart_2019_sparse-tensor,Alex_2020_safe-tensor} implementing tensor algebra.
\begin{itemize}
  \item first order (constants $e_{38},e_{39},e_{40}$):
\begin{equation}
  \label{aA}
    a_A^{\hphantom AI} H^A_{\hphantom AI} = \Big\lbrack e_{38} \cdot \eta_{ac}\eta_{bd} \eta_{pq} + e_{39} \cdot \eta_{ac} \eta_{bp} \eta_{dq} + e_{40} \cdot \epsilon_{abcd}\eta_{pq}\Big\rbrack \times \eta^{pr} \eta^{qs} H^{abcd}_{\hphantom{abcd},rs}
\end{equation}
\item second order (constants $e_{1},\dots,e_{37}$):
\begin{equation}
  \label{aAB}
  \begin{aligned}
    & a_{AB} H^A H^B = \Big\lbrack \\
    & \hphantom{{}+{}} e_{1} \cdot \eta_{a c} \eta_{b d} \eta_{e g} \eta_{f h} + e_{2} \cdot \eta_{a c} \eta_{b e} \eta_{d g} \eta_{f h} + e_{3} \cdot \eta_{a e} \eta_{b f} \eta_{c g} \eta_{d h} \\
    & + e_{4} \cdot \eta_{a e} \eta_{b g} \eta_{c f} \eta_{d h} + e_{5} \cdot \epsilon_{a b c d} \eta_{e g} \eta_{f h} + e_{6} \cdot \epsilon_{a b e f} \eta_{c g} \eta_{d h} \\
    & \Big\rbrack \times H^{abcd} H^{efgh}
  \end{aligned}
\end{equation}
\begin{equation}
  \label{aApBq}
  \begin{aligned}
    & a_{A\hphantom{p}B}^{\hphantom Ap\hphantom Bq} H^A_{\hphantom Ap} H^B_{\hphantom Bq} = \Big\lbrack \\
    & \hphantom{{}+{}} e_{7} \cdot \eta_{a c} \eta_{b d} \eta_{p e} \eta_{f g} \eta_{h q} + e_{8} \cdot \eta_{a c} \eta_{b d} \eta_{p q} \eta_{e g} \eta_{f h} + e_{9} \cdot \eta_{a c} \eta_{b p} \eta_{d e} \eta_{f g} \eta_{h q} \\
    & + e_{10} \cdot \eta_{a c} \eta_{b e} \eta_{d g} \eta_{p f} \eta_{h q} + e_{11} \cdot \eta_{a c} \eta_{b e} \eta_{d g} \eta_{p q} \eta_{f h} + e_{12} \cdot \eta_{a c} \eta_{b e} \eta_{d q} \eta_{p g} \eta_{f h} \\
    & + e_{13} \cdot \eta_{a p} \eta_{b e} \eta_{c f} \eta_{d g} \eta_{h q} + e_{14} \cdot \eta_{a p} \eta_{b e} \eta_{c g} \eta_{d h} \eta_{f q} + e_{15} \cdot \eta_{a e} \eta_{b f} \eta_{c g} \eta_{d h} \eta_{p q} \\
    & + e_{16} \cdot \epsilon_{a b c d} \eta_{p e} \eta_{f g} \eta_{h q} + e_{17} \cdot \epsilon_{a b c d} \eta_{p q} \eta_{e g} \eta_{f h} + e_{18} \cdot \epsilon_{a b p e} \eta_{c f} \eta_{d g} \eta_{h q} \\
    & + e_{19} \cdot \epsilon_{a b p e} \eta_{c g} \eta_{d q} \eta_{f h} + e_{20} \cdot \epsilon_{a b e f} \eta_{c p} \eta_{d g} \eta_{h q} + e_{21} \cdot \epsilon_{a b e f} \eta_{c g} \eta_{d h} \eta_{p q} \\
    & \Big\rbrack \times \eta^{pr} \eta^{qs} H^{abcd}_{\hphantom{abcd},r} H^{efgh}_{\hphantom{efgh},s}
  \end{aligned}
\end{equation}
\begin{equation}
  \label{aABI}
  \begin{aligned}
    & a_{AB}^{\hphantom{AB}I} H^A H^B_{\hphantom BI} = \Big\lbrack \\
    & \hphantom{{}+{}} e_{22} \cdot \eta_{a c} \eta_{b d} \eta_{e g} \eta_{f h} \eta_{p q} + e_{23} \cdot \eta_{a c} \eta_{b d} \eta_{e g} \eta_{f p} \eta_{h q} + e_{24} \cdot \eta_{a c} \eta_{b e} \eta_{d g} \eta_{f h} \eta_{p q} \\
    & + e_{25} \cdot \eta_{a c} \eta_{b e} \eta_{d g} \eta_{f p} \eta_{h q} + e_{26} \cdot \eta_{a c} \eta_{b e} \eta_{d p} \eta_{f g} \eta_{h q} + e_{27} \cdot \eta_{a c} \eta_{b p} \eta_{d q} \eta_{e g} \eta_{f h} \\
    & + e_{28} \cdot \eta_{a e} \eta_{b f} \eta_{c g} \eta_{d h} \eta_{p q} + e_{29} \cdot \eta_{a e} \eta_{b f} \eta_{c g} \eta_{d p} \eta_{h q} + e_{30} \cdot \eta_{a e} \eta_{b g} \eta_{c f} \eta_{d h} \eta_{p q} \\
    & + e_{31} \cdot \epsilon_{a b c d} \eta_{e g} \eta_{f h} \eta_{p q} + e_{32} \cdot \epsilon_{a b c d} \eta_{e g} \eta_{f p} \eta_{h q} + e_{33} \cdot \epsilon_{a b e f} \eta_{c g} \eta_{d h} \eta_{p q} \\
    & + e_{34} \cdot \epsilon_{a b e f} \eta_{c g} \eta_{d p} \eta_{h q} + e_{35} \cdot \epsilon_{a b e p} \eta_{c f} \eta_{d g} \eta_{h q} + e_{36} \cdot \epsilon_{a b e p} \eta_{c g} \eta_{d h} \eta_{f q} \\
    & + e_{37} \cdot \epsilon_{e f g h} \eta_{a c} \eta_{b d} \eta_{p q} \Big\rbrack \times \eta^{pr} \eta^{qs} H^{abcd} H^{efgh}_{\hphantom{efgh},rs}
  \end{aligned}
\end{equation}
\item third order (constants $e_{41},\dots,e_{237}$):
\begin{equation}
  \label{aABC}
  \begin{aligned}
    a_{ABC} H^A H^B H^C = \Big\lbrack & \hphantom{{}+{}} e_{41} \cdot \eta_{a c} \eta_{b d} \eta_{e g} \eta_{f h} \eta_{i k} \eta_{j l} + e_{42} \cdot \eta_{a c} \eta_{b d} \eta_{e g} \eta_{f i} \eta_{h k} \eta_{j l} \\
    & + e_{43} \cdot \eta_{a c} \eta_{b d} \eta_{e i} \eta_{f j} \eta_{g k} \eta_{h l} + e_{44} \cdot \eta_{a c} \eta_{b d} \eta_{e i} \eta_{f k} \eta_{g j} \eta_{h l} \\
    & + e_{45} \cdot \eta_{a c} \eta_{b e} \eta_{d g} \eta_{f i} \eta_{h k} \eta_{j l} + e_{46} \cdot \eta_{a c} \eta_{b e} \eta_{d i} \eta_{f g} \eta_{h k} \eta_{j l} \\
    & + e_{47} \cdot \eta_{a e} \eta_{b f} \eta_{c i} \eta_{d j} \eta_{g k} \eta_{h l} + e_{48} \cdot \eta_{a e} \eta_{b f} \eta_{c i} \eta_{d k} \eta_{g j} \eta_{h l} \\
    & + e_{49} \cdot \epsilon_{a b c d} \eta_{e g} \eta_{f h} \eta_{i k} \eta_{j l} + e_{50} \cdot \epsilon_{a b c d} \eta_{e g} \eta_{f i} \eta_{h k} \eta_{j l} \\
    & + e_{51} \cdot \epsilon_{a b c d} \eta_{e i} \eta_{f j} \eta_{g k} \eta_{h l} + e_{52} \cdot \epsilon_{a b c d} \eta_{e i} \eta_{f k} \eta_{g j} \eta_{h l} \\
    & + e_{53} \cdot \epsilon_{a b e f} \eta_{c g} \eta_{d h} \eta_{i k} \eta_{j l} + e_{54} \cdot \epsilon_{a b e f} \eta_{c g} \eta_{d i} \eta_{h k} \eta_{j l} \\
    & + e_{55} \cdot \epsilon_{a b e f} \eta_{c i} \eta_{d j} \eta_{g k} \eta_{h l} \Big\rbrack \times H^{abcd} H^{efgh} H^{ijkl}
  \end{aligned}
\end{equation}
\begingroup
\allowdisplaybreaks
\begin{align} \label{aABpCq}
    & a_{AB\hphantom pC}^{\hphantom{AB}p\hphantom Cq} H^A H^B_{\hphantom Bp} H^C_{\hphantom Cq} = \Big\lbrack \nonumber \\
    & \hphantom{{}+{}} e_{56} \cdot \eta_{a c} \eta_{b d} \eta_{e g} \eta_{f h} \eta_{p i} \eta_{j k} \eta_{l q} + e_{57} \cdot \eta_{a c} \eta_{b d} \eta_{e g} \eta_{f h} \eta_{p q} \eta_{i k} \eta_{j l} + e_{58} \cdot \eta_{a c} \eta_{b d} \eta_{e g} \eta_{f p} \eta_{h i} \eta_{j k} \eta_{l q} \nonumber \\
    & + e_{59} \cdot \eta_{a c} \eta_{b d} \eta_{e g} \eta_{f i} \eta_{h k} \eta_{p j} \eta_{l q} + e_{60} \cdot \eta_{a c} \eta_{b d} \eta_{e g} \eta_{f i} \eta_{h k} \eta_{p q} \eta_{j l} + e_{61} \cdot \eta_{a c} \eta_{b d} \eta_{e g} \eta_{f i} \eta_{h q} \eta_{p k} \eta_{j l} \nonumber \\
    & + e_{62} \cdot \eta_{a c} \eta_{b d} \eta_{e p} \eta_{f i} \eta_{g j} \eta_{h k} \eta_{l q} + e_{63} \cdot \eta_{a c} \eta_{b d} \eta_{e p} \eta_{f i} \eta_{g k} \eta_{h l} \eta_{j q} + e_{64} \cdot \eta_{a c} \eta_{b d} \eta_{e i} \eta_{f j} \eta_{g k} \eta_{h l} \eta_{p q} \nonumber \\
    & + e_{65} \cdot \eta_{a c} \eta_{b e} \eta_{d g} \eta_{f h} \eta_{p i} \eta_{j k} \eta_{l q} + e_{66} \cdot \eta_{a c} \eta_{b e} \eta_{d g} \eta_{f h} \eta_{p q} \eta_{i k} \eta_{j l} + e_{67} \cdot \eta_{a c} \eta_{b e} \eta_{d g} \eta_{f p} \eta_{h i} \eta_{j k} \eta_{l q} \nonumber \\
    & + e_{68} \cdot \eta_{a c} \eta_{b e} \eta_{d g} \eta_{f p} \eta_{h q} \eta_{i k} \eta_{j l} + e_{69} \cdot \eta_{a c} \eta_{b e} \eta_{d g} \eta_{f i} \eta_{h k} \eta_{p j} \eta_{l q} + e_{70} \cdot \eta_{a c} \eta_{b e} \eta_{d g} \eta_{f i} \eta_{h k} \eta_{p q} \eta_{j l} \nonumber \\
    & + e_{71} \cdot \eta_{a c} \eta_{b e} \eta_{d g} \eta_{f i} \eta_{h q} \eta_{p k} \eta_{j l} + e_{72} \cdot \eta_{a c} \eta_{b e} \eta_{d p} \eta_{f g} \eta_{h i} \eta_{j k} \eta_{l q} + e_{73} \cdot \eta_{a c} \eta_{b e} \eta_{d p} \eta_{f g} \eta_{h q} \eta_{i k} \eta_{j l} \nonumber \\
    & + e_{74} \cdot \eta_{a c} \eta_{b e} \eta_{d p} \eta_{f i} \eta_{g j} \eta_{h k} \eta_{l q} + e_{75} \cdot \eta_{a c} \eta_{b e} \eta_{d p} \eta_{f i} \eta_{g k} \eta_{h l} \eta_{j q} + e_{76} \cdot \eta_{a c} \eta_{b e} \eta_{d p} \eta_{f i} \eta_{g k} \eta_{h q} \eta_{j l} \nonumber \\
    & + e_{77} \cdot \eta_{a c} \eta_{b e} \eta_{d i} \eta_{f g} \eta_{h p} \eta_{j k} \eta_{l q} + e_{78} \cdot \eta_{a c} \eta_{b e} \eta_{d i} \eta_{f g} \eta_{h j} \eta_{p k} \eta_{l q} + e_{79} \cdot \eta_{a c} \eta_{b e} \eta_{d i} \eta_{f g} \eta_{h k} \eta_{p j} \eta_{l q} \nonumber \\
    & + e_{80} \cdot \eta_{a c} \eta_{b e} \eta_{d i} \eta_{f g} \eta_{h k} \eta_{p l} \eta_{j q} + e_{81} \cdot \eta_{a c} \eta_{b e} \eta_{d i} \eta_{f g} \eta_{h k} \eta_{p q} \eta_{j l} + e_{82} \cdot \eta_{a c} \eta_{b e} \eta_{d i} \eta_{f g} \eta_{h q} \eta_{p k} \eta_{j l} \nonumber \\
    & + e_{83} \cdot \eta_{a c} \eta_{b e} \eta_{d i} \eta_{f p} \eta_{g j} \eta_{h k} \eta_{l q} + e_{84} \cdot \eta_{a c} \eta_{b e} \eta_{d i} \eta_{f p} \eta_{g k} \eta_{h l} \eta_{j q} + e_{85} \cdot \eta_{a c} \eta_{b e} \eta_{d i} \eta_{f j} \eta_{g p} \eta_{h k} \eta_{l q} \nonumber \\
    & + e_{86} \cdot \eta_{a c} \eta_{b e} \eta_{d i} \eta_{f k} \eta_{g j} \eta_{h q} \eta_{p l} + e_{87} \cdot \eta_{a c} \eta_{b e} \eta_{d q} \eta_{f g} \eta_{h p} \eta_{i k} \eta_{j l} + e_{88} \cdot \eta_{a c} \eta_{b e} \eta_{d q} \eta_{f g} \eta_{h i} \eta_{p k} \eta_{j l} \nonumber \\
    & + e_{89} \cdot \eta_{a c} \eta_{b p} \eta_{d q} \eta_{e g} \eta_{f h} \eta_{i k} \eta_{j l} + e_{90} \cdot \eta_{a e} \eta_{b f} \eta_{c g} \eta_{d h} \eta_{p i} \eta_{j k} \eta_{l q} + e_{91} \cdot \eta_{a e} \eta_{b f} \eta_{c g} \eta_{d h} \eta_{p q} \eta_{i k} \eta_{j l} \nonumber \\
    & + e_{92} \cdot \eta_{a e} \eta_{b f} \eta_{c g} \eta_{d p} \eta_{h i} \eta_{j k} \eta_{l q} + e_{93} \cdot \eta_{a e} \eta_{b f} \eta_{c g} \eta_{d p} \eta_{h q} \eta_{i k} \eta_{j l} + e_{94} \cdot \eta_{a e} \eta_{b f} \eta_{c g} \eta_{d i} \eta_{h p} \eta_{j k} \eta_{l q} \nonumber \\
    & + e_{95} \cdot \eta_{a e} \eta_{b f} \eta_{c g} \eta_{d i} \eta_{h j} \eta_{p k} \eta_{l q} + e_{96} \cdot \eta_{a e} \eta_{b f} \eta_{c g} \eta_{d i} \eta_{h k} \eta_{p j} \eta_{l q} + e_{97} \cdot \eta_{a e} \eta_{b f} \eta_{c g} \eta_{d i} \eta_{h k} \eta_{p l} \eta_{j q} \nonumber \\
    & + e_{98} \cdot \eta_{a e} \eta_{b f} \eta_{c g} \eta_{d i} \eta_{h k} \eta_{p q} \eta_{j l} + e_{99} \cdot \eta_{a e} \eta_{b f} \eta_{c g} \eta_{d i} \eta_{h q} \eta_{p k} \eta_{j l} + e_{100} \cdot \eta_{a e} \eta_{b f} \eta_{c g} \eta_{d q} \eta_{h p} \eta_{i k} \eta_{j l} \nonumber \\
    & + e_{101} \cdot \eta_{a e} \eta_{b f} \eta_{c g} \eta_{d q} \eta_{h i} \eta_{p k} \eta_{j l} + e_{102} \cdot \eta_{a e} \eta_{b f} \eta_{c p} \eta_{d i} \eta_{g j} \eta_{h k} \eta_{l q} + e_{103} \cdot \eta_{a e} \eta_{b f} \eta_{c p} \eta_{d i} \eta_{g k} \eta_{h l} \eta_{j q} \nonumber \\
    & + e_{104} \cdot \eta_{a e} \eta_{b f} \eta_{c p} \eta_{d i} \eta_{g k} \eta_{h q} \eta_{j l} + e_{105} \cdot \eta_{a e} \eta_{b f} \eta_{c i} \eta_{d j} \eta_{g p} \eta_{h k} \eta_{l q} + e_{106} \cdot \eta_{a e} \eta_{b f} \eta_{c i} \eta_{d j} \eta_{g k} \eta_{h l} \eta_{p q} \nonumber \\
    & + e_{107} \cdot \eta_{a e} \eta_{b f} \eta_{c i} \eta_{d k} \eta_{g p} \eta_{h j} \eta_{l q} + e_{108} \cdot \eta_{a e} \eta_{b g} \eta_{c f} \eta_{d h} \eta_{p i} \eta_{j k} \eta_{l q} + e_{109} \cdot \eta_{a e} \eta_{b g} \eta_{c f} \eta_{d h} \eta_{p q} \eta_{i k} \eta_{j l} \nonumber \\
    & + e_{110} \cdot \eta_{a e} \eta_{b p} \eta_{c f} \eta_{d i} \eta_{g j} \eta_{h k} \eta_{l q} + e_{111} \cdot \eta_{a e} \eta_{b p} \eta_{c f} \eta_{d i} \eta_{g k} \eta_{h q} \eta_{j l} + e_{112} \cdot \eta_{a e} \eta_{b i} \eta_{c f} \eta_{d j} \eta_{g p} \eta_{h k} \eta_{l q} \nonumber \\
    & + e_{113} \cdot \eta_{a e} \eta_{b i} \eta_{c f} \eta_{d j} \eta_{g k} \eta_{h l} \eta_{p q} + e_{114} \cdot \epsilon_{a b c d} \eta_{e g} \eta_{f h} \eta_{p i} \eta_{j k} \eta_{l q} + e_{115} \cdot \epsilon_{a b c d} \eta_{e g} \eta_{f h} \eta_{p q} \eta_{i k} \eta_{j l} \nonumber \\
    & + e_{116} \cdot \epsilon_{a b c d} \eta_{e g} \eta_{f p} \eta_{h i} \eta_{j k} \eta_{l q} + e_{117} \cdot \epsilon_{a b c d} \eta_{e g} \eta_{f i} \eta_{h k} \eta_{p j} \eta_{l q} + e_{118} \cdot \epsilon_{a b c d} \eta_{e g} \eta_{f i} \eta_{h k} \eta_{p q} \eta_{j l} \nonumber \\
    & + e_{119} \cdot \epsilon_{a b c d} \eta_{e g} \eta_{f i} \eta_{h q} \eta_{p k} \eta_{j l} + e_{120} \cdot \epsilon_{a b c d} \eta_{e p} \eta_{f i} \eta_{g j} \eta_{h k} \eta_{l q} + e_{121} \cdot \epsilon_{a b c d} \eta_{e p} \eta_{f i} \eta_{g k} \eta_{h l} \eta_{j q} \nonumber \\
    & + e_{122} \cdot \epsilon_{a b c d} \eta_{e i} \eta_{f j} \eta_{g k} \eta_{h l} \eta_{p q} + e_{123} \cdot \epsilon_{a b e f} \eta_{c g} \eta_{d h} \eta_{p i} \eta_{j k} \eta_{l q} + e_{124} \cdot \epsilon_{a b e f} \eta_{c g} \eta_{d h} \eta_{p q} \eta_{i k} \eta_{j l} \nonumber \\
    & + e_{125} \cdot \epsilon_{a b e f} \eta_{c g} \eta_{d p} \eta_{h i} \eta_{j k} \eta_{l q} + e_{126} \cdot \epsilon_{a b e f} \eta_{c g} \eta_{d p} \eta_{h q} \eta_{i k} \eta_{j l} + e_{127} \cdot \epsilon_{a b e f} \eta_{c g} \eta_{d i} \eta_{h p} \eta_{j k} \eta_{l q} \nonumber \\
    & + e_{128} \cdot \epsilon_{a b e f} \eta_{c g} \eta_{d i} \eta_{h j} \eta_{p k} \eta_{l q} + e_{129} \cdot \epsilon_{a b e f} \eta_{c g} \eta_{d i} \eta_{h k} \eta_{p j} \eta_{l q} + e_{130} \cdot \epsilon_{a b e f} \eta_{c g} \eta_{d i} \eta_{h k} \eta_{p l} \eta_{j q} \nonumber \\
    & + e_{131} \cdot \epsilon_{a b e f} \eta_{c g} \eta_{d i} \eta_{h k} \eta_{p q} \eta_{j l} + e_{132} \cdot \epsilon_{a b e f} \eta_{c g} \eta_{d i} \eta_{h q} \eta_{p k} \eta_{j l} + e_{133} \cdot \epsilon_{a b e f} \eta_{c g} \eta_{d q} \eta_{h p} \eta_{i k} \eta_{j l} \nonumber \\
    & + e_{134} \cdot \epsilon_{a b e f} \eta_{c g} \eta_{d q} \eta_{h i} \eta_{p k} \eta_{j l} + e_{135} \cdot \epsilon_{a b e f} \eta_{c p} \eta_{d i} \eta_{g j} \eta_{h k} \eta_{l q} + e_{136} \cdot \epsilon_{a b e f} \eta_{c p} \eta_{d i} \eta_{g k} \eta_{h l} \eta_{j q} \nonumber \\
    & + e_{137} \cdot \epsilon_{a b e f} \eta_{c p} \eta_{d i} \eta_{g k} \eta_{h q} \eta_{j l} + e_{138} \cdot \epsilon_{a b e f} \eta_{c i} \eta_{d j} \eta_{g p} \eta_{h k} \eta_{l q} + e_{139} \cdot \epsilon_{a b e f} \eta_{c i} \eta_{d j} \eta_{g k} \eta_{h l} \eta_{p q} \nonumber \\
    & + e_{140} \cdot \epsilon_{a b e f} \eta_{c i} \eta_{d k} \eta_{g p} \eta_{h j} \eta_{l q} + e_{141} \cdot \epsilon_{a b e p} \eta_{c f} \eta_{d g} \eta_{h i} \eta_{j k} \eta_{l q} + e_{142} \cdot \epsilon_{a b e p} \eta_{c f} \eta_{d g} \eta_{h q} \eta_{i k} \eta_{j l} \nonumber \\
    & + e_{143} \cdot \epsilon_{a b e p} \eta_{c f} \eta_{d i} \eta_{g j} \eta_{h k} \eta_{l q} + e_{144} \cdot \epsilon_{a b e p} \eta_{c f} \eta_{d i} \eta_{g k} \eta_{h l} \eta_{j q} + e_{145} \cdot \epsilon_{a b e p} \eta_{c f} \eta_{d i} \eta_{g k} \eta_{h q} \eta_{j l} \nonumber \\
    & + e_{146} \cdot \epsilon_{a b e p} \eta_{c g} \eta_{d h} \eta_{f i} \eta_{j k} \eta_{l q} + e_{147} \cdot \epsilon_{a b e p} \eta_{c g} \eta_{d h} \eta_{f q} \eta_{i k} \eta_{j l} + e_{148} \cdot \epsilon_{a b e p} \eta_{c g} \eta_{d i} \eta_{f j} \eta_{h k} \eta_{l q} \nonumber \\
    & + e_{149} \cdot \epsilon_{a b e p} \eta_{c g} \eta_{d i} \eta_{f k} \eta_{h q} \eta_{j l} + e_{150} \cdot \epsilon_{a b e i} \eta_{c f} \eta_{d g} \eta_{h p} \eta_{j k} \eta_{l q} + e_{151} \cdot \epsilon_{a b e i} \eta_{c f} \eta_{d g} \eta_{h j} \eta_{p k} \eta_{l q} \nonumber \\
    & + e_{152} \cdot \epsilon_{a b e i} \eta_{c f} \eta_{d g} \eta_{h k} \eta_{p j} \eta_{l q} + e_{153} \cdot \epsilon_{a b e i} \eta_{c f} \eta_{d g} \eta_{h k} \eta_{p l} \eta_{j q} + e_{154} \cdot \epsilon_{a b e i} \eta_{c f} \eta_{d g} \eta_{h k} \eta_{p q} \eta_{j l} \nonumber \\
    & + e_{155} \cdot \epsilon_{a b e i} \eta_{c f} \eta_{d g} \eta_{h q} \eta_{p k} \eta_{j l} + e_{156} \cdot \epsilon_{a b e i} \eta_{c f} \eta_{d p} \eta_{g j} \eta_{h k} \eta_{l q} + e_{157} \cdot \epsilon_{a b e i} \eta_{c f} \eta_{d p} \eta_{g k} \eta_{h l} \eta_{j q} \nonumber \\
    & + e_{158} \cdot \epsilon_{a b e i} \eta_{c f} \eta_{d p} \eta_{g k} \eta_{h q} \eta_{j l} + e_{159} \cdot \epsilon_{a b e i} \eta_{c f} \eta_{d j} \eta_{g p} \eta_{h k} \eta_{l q} + e_{160} \cdot \epsilon_{a b e i} \eta_{c f} \eta_{d j} \eta_{g k} \eta_{h l} \eta_{p q} \nonumber \\
    & + e_{161} \cdot \epsilon_{a b e i} \eta_{c f} \eta_{d k} \eta_{g p} \eta_{h j} \eta_{l q} + e_{162} \cdot \epsilon_{a b e i} \eta_{c f} \eta_{d k} \eta_{g p} \eta_{h l} \eta_{j q} + e_{163} \cdot \epsilon_{a b e i} \eta_{c f} \eta_{d k} \eta_{g j} \eta_{h q} \eta_{p l} \nonumber \\
    & + e_{164} \cdot \epsilon_{a b e q} \eta_{c f} \eta_{d g} \eta_{h p} \eta_{i k} \eta_{j l} + e_{165} \cdot \epsilon_{e f g h} \eta_{a c} \eta_{b d} \eta_{p i} \eta_{j k} \eta_{l q} \nonumber \\
    & \Big\rbrack \times \eta^{pr} \eta^{qs} H^{abcd} H^{efgh}_{\hphantom{efgh},r} H^{ijkl}_{\hphantom{ijkl},s}
\end{align}
\endgroup
\begingroup
\allowdisplaybreaks
\begin{align}
  \label{aABCI}
    & a_{ABC}^{\hphantom{ABC}I} H^A H^B H^C_{\hphantom CI} = \Big\lbrack \nonumber \\
    & \hphantom{{}+{}} e_{166} \cdot \eta_{a c} \eta_{b d} \eta_{e g} \eta_{f h} \eta_{i k} \eta_{j l} \eta_{p q} + e_{167} \cdot \eta_{a c} \eta_{b d} \eta_{e g} \eta_{f h} \eta_{i k} \eta_{j p} \eta_{l q} + e_{168} \cdot \eta_{a c} \eta_{b d} \eta_{e g} \eta_{f i} \eta_{h k} \eta_{j l} \eta_{p q} \nonumber \\
    & + e_{169} \cdot \eta_{a c} \eta_{b d} \eta_{e g} \eta_{f i} \eta_{h k} \eta_{j p} \eta_{l q} + e_{170} \cdot \eta_{a c} \eta_{b d} \eta_{e g} \eta_{f i} \eta_{h p} \eta_{j k} \eta_{l q} + e_{171} \cdot \eta_{a c} \eta_{b d} \eta_{e g} \eta_{f p} \eta_{h q} \eta_{i k} \eta_{j l} \nonumber \\
    & + e_{172} \cdot \eta_{a c} \eta_{b d} \eta_{e i} \eta_{f j} \eta_{g k} \eta_{h l} \eta_{p q} + e_{173} \cdot \eta_{a c} \eta_{b d} \eta_{e i} \eta_{f j} \eta_{g k} \eta_{h p} \eta_{l q} + e_{174} \cdot \eta_{a c} \eta_{b d} \eta_{e i} \eta_{f k} \eta_{g j} \eta_{h l} \eta_{p q} \nonumber \\
    & + e_{175} \cdot \eta_{a c} \eta_{b e} \eta_{d g} \eta_{f h} \eta_{i k} \eta_{j l} \eta_{p q} + e_{176} \cdot \eta_{a c} \eta_{b e} \eta_{d g} \eta_{f h} \eta_{i k} \eta_{j p} \eta_{l q} + e_{177} \cdot \eta_{a c} \eta_{b e} \eta_{d g} \eta_{f i} \eta_{h k} \eta_{j l} \eta_{p q} \nonumber \\
    & + e_{178} \cdot \eta_{a c} \eta_{b e} \eta_{d g} \eta_{f i} \eta_{h k} \eta_{j p} \eta_{l q} + e_{179} \cdot \eta_{a c} \eta_{b e} \eta_{d g} \eta_{f i} \eta_{h p} \eta_{j k} \eta_{l q} + e_{180} \cdot \eta_{a c} \eta_{b e} \eta_{d g} \eta_{f p} \eta_{h q} \eta_{i k} \eta_{j l} \nonumber \\
    & + e_{181} \cdot \eta_{a c} \eta_{b e} \eta_{d i} \eta_{f g} \eta_{h k} \eta_{j l} \eta_{p q} + e_{182} \cdot \eta_{a c} \eta_{b e} \eta_{d i} \eta_{f g} \eta_{h k} \eta_{j p} \eta_{l q} + e_{183} \cdot \eta_{a c} \eta_{b e} \eta_{d i} \eta_{f g} \eta_{h p} \eta_{j k} \eta_{l q} \nonumber \\
    & + e_{184} \cdot \eta_{a c} \eta_{b e} \eta_{d i} \eta_{f j} \eta_{g k} \eta_{h l} \eta_{p q} + e_{185} \cdot \eta_{a c} \eta_{b e} \eta_{d i} \eta_{f j} \eta_{g k} \eta_{h p} \eta_{l q} + e_{186} \cdot \eta_{a c} \eta_{b e} \eta_{d i} \eta_{f k} \eta_{g j} \eta_{h p} \eta_{l q} \nonumber \\
    & + e_{187} \cdot \eta_{a c} \eta_{b e} \eta_{d i} \eta_{f k} \eta_{g l} \eta_{h p} \eta_{j q} + e_{188} \cdot \eta_{a c} \eta_{b e} \eta_{d i} \eta_{f p} \eta_{g j} \eta_{h k} \eta_{l q} + e_{189} \cdot \eta_{a c} \eta_{b e} \eta_{d i} \eta_{f p} \eta_{g k} \eta_{h l} \eta_{j q} \nonumber \\
    & + e_{190} \cdot \eta_{a c} \eta_{b e} \eta_{d i} \eta_{f p} \eta_{g k} \eta_{h q} \eta_{j l} + e_{191} \cdot \eta_{a c} \eta_{b e} \eta_{d p} \eta_{f g} \eta_{h q} \eta_{i k} \eta_{j l} + e_{192} \cdot \eta_{a c} \eta_{b e} \eta_{d p} \eta_{f i} \eta_{g j} \eta_{h k} \eta_{l q} \nonumber \\
    & + e_{193} \cdot \eta_{a c} \eta_{b i} \eta_{d k} \eta_{e g} \eta_{f p} \eta_{h q} \eta_{j l} + e_{194} \cdot \eta_{a c} \eta_{b i} \eta_{d k} \eta_{e j} \eta_{f p} \eta_{g l} \eta_{h q} + e_{195} \cdot \eta_{a e} \eta_{b f} \eta_{c g} \eta_{d h} \eta_{i k} \eta_{j l} \eta_{p q} \nonumber \\
    & + e_{196} \cdot \eta_{a e} \eta_{b f} \eta_{c g} \eta_{d h} \eta_{i k} \eta_{j p} \eta_{l q} + e_{197} \cdot \eta_{a e} \eta_{b f} \eta_{c i} \eta_{d j} \eta_{g k} \eta_{h l} \eta_{p q} + e_{198} \cdot \eta_{a e} \eta_{b f} \eta_{c i} \eta_{d j} \eta_{g k} \eta_{h p} \eta_{l q} \nonumber \\
    & + e_{199} \cdot \eta_{a e} \eta_{b f} \eta_{c i} \eta_{d k} \eta_{g j} \eta_{h l} \eta_{p q} + e_{200} \cdot \eta_{a e} \eta_{b g} \eta_{c f} \eta_{d h} \eta_{i k} \eta_{j l} \eta_{p q} + e_{201} \cdot \eta_{a e} \eta_{b g} \eta_{c f} \eta_{d h} \eta_{i k} \eta_{j p} \eta_{l q} \nonumber \\
    & + e_{202} \cdot \eta_{a e} \eta_{b g} \eta_{c i} \eta_{d j} \eta_{f k} \eta_{h l} \eta_{p q} + e_{203} \cdot \eta_{a e} \eta_{b g} \eta_{c i} \eta_{d j} \eta_{f k} \eta_{h p} \eta_{l q} + e_{204} \cdot \eta_{a e} \eta_{b i} \eta_{c g} \eta_{d k} \eta_{f p} \eta_{h q} \eta_{j l} \nonumber \\
    & + e_{205} \cdot \epsilon_{a b c d} \eta_{e g} \eta_{f h} \eta_{i k} \eta_{j l} \eta_{p q} + e_{206} \cdot \epsilon_{a b c d} \eta_{e g} \eta_{f h} \eta_{i k} \eta_{j p} \eta_{l q} + e_{207} \cdot \epsilon_{a b c d} \eta_{e g} \eta_{f i} \eta_{h k} \eta_{j l} \eta_{p q} \nonumber \\
    & + e_{208} \cdot \epsilon_{a b c d} \eta_{e g} \eta_{f i} \eta_{h k} \eta_{j p} \eta_{l q} + e_{209} \cdot \epsilon_{a b c d} \eta_{e g} \eta_{f i} \eta_{h p} \eta_{j k} \eta_{l q} + e_{210} \cdot \epsilon_{a b c d} \eta_{e g} \eta_{f p} \eta_{h q} \eta_{i k} \eta_{j l} \nonumber \\
    & + e_{211} \cdot \epsilon_{a b c d} \eta_{e i} \eta_{f j} \eta_{g k} \eta_{h l} \eta_{p q} + e_{212} \cdot \epsilon_{a b c d} \eta_{e i} \eta_{f j} \eta_{g k} \eta_{h p} \eta_{l q} + e_{213} \cdot \epsilon_{a b c d} \eta_{e i} \eta_{f k} \eta_{g j} \eta_{h l} \eta_{p q} \nonumber \\
    & + e_{214} \cdot \epsilon_{a b e f} \eta_{c g} \eta_{d h} \eta_{i k} \eta_{j l} \eta_{p q} + e_{215} \cdot \epsilon_{a b e f} \eta_{c g} \eta_{d h} \eta_{i k} \eta_{j p} \eta_{l q} + e_{216} \cdot \epsilon_{a b e f} \eta_{c g} \eta_{d i} \eta_{h k} \eta_{j l} \eta_{p q} \nonumber \\
    & + e_{217} \cdot \epsilon_{a b e f} \eta_{c g} \eta_{d i} \eta_{h k} \eta_{j p} \eta_{l q} + e_{218} \cdot \epsilon_{a b e f} \eta_{c g} \eta_{d i} \eta_{h p} \eta_{j k} \eta_{l q} + e_{219} \cdot \epsilon_{a b e f} \eta_{c g} \eta_{d p} \eta_{h q} \eta_{i k} \eta_{j l} \nonumber \\
    & + e_{220} \cdot \epsilon_{a b e f} \eta_{c i} \eta_{d j} \eta_{g k} \eta_{h l} \eta_{p q} + e_{221} \cdot \epsilon_{a b e f} \eta_{c i} \eta_{d j} \eta_{g k} \eta_{h p} \eta_{l q} + e_{222} \cdot \epsilon_{a b e f} \eta_{c i} \eta_{d k} \eta_{g j} \eta_{h l} \eta_{p q} \nonumber \\
    & + e_{223} \cdot \epsilon_{a b e i} \eta_{c f} \eta_{d j} \eta_{g k} \eta_{h l} \eta_{p q} + e_{224} \cdot \epsilon_{a b e i} \eta_{c f} \eta_{d j} \eta_{g k} \eta_{h p} \eta_{l q} + e_{225} \cdot \epsilon_{a b e i} \eta_{c f} \eta_{d k} \eta_{g j} \eta_{h l} \eta_{p q} \nonumber \\
    & + e_{226} \cdot \epsilon_{a b e i} \eta_{c f} \eta_{d k} \eta_{g j} \eta_{h p} \eta_{l q} + e_{227} \cdot \epsilon_{a b e i} \eta_{c f} \eta_{d k} \eta_{g l} \eta_{h p} \eta_{j q} + e_{228} \cdot \epsilon_{a b e i} \eta_{c f} \eta_{d p} \eta_{g j} \eta_{h k} \eta_{l q} \nonumber \\
    & + e_{229} \cdot \epsilon_{a b e i} \eta_{c f} \eta_{d p} \eta_{g k} \eta_{h l} \eta_{j q} + e_{230} \cdot \epsilon_{a b e i} \eta_{c f} \eta_{d p} \eta_{g k} \eta_{h q} \eta_{j l} + e_{231} \cdot \epsilon_{a b e p} \eta_{c f} \eta_{d i} \eta_{g j} \eta_{h k} \eta_{l q} \nonumber \\
    & + e_{232} \cdot \epsilon_{a b i j} \eta_{c e} \eta_{d f} \eta_{g k} \eta_{h l} \eta_{p q} + e_{233} \cdot \epsilon_{a b i j} \eta_{c e} \eta_{d f} \eta_{g k} \eta_{h p} \eta_{l q} + e_{234} \cdot \epsilon_{a b i j} \eta_{c e} \eta_{d k} \eta_{f p} \eta_{g l} \eta_{h q} \nonumber \\
    & + e_{235} \cdot \epsilon_{a b i p} \eta_{c e} \eta_{d f} \eta_{g j} \eta_{h k} \eta_{l q} + e_{236} \cdot \epsilon_{a b i p} \eta_{c e} \eta_{d f} \eta_{g k} \eta_{h l} \eta_{j q} + e_{237} \cdot \epsilon_{i j k l} \eta_{a c} \eta_{b d} \eta_{e g} \eta_{f h} \eta_{p q} \nonumber \\
    & \Big\rbrack \times \eta^{pr} \eta^{qs} H^{abcd} H^{efgh} H^{ijkl}_{\hphantom{ijkl},rs}
\end{align}
\endgroup
\end{itemize}
\section{Reduction}
\label{reduction}

We used the Haskell library \cite{Alex_2020_safe-tensor} in order to solve the previously obtained ansätze for diffeomorphism invariance (see also \cite{Alex_2020}). The source code and the solution are publicly available \cite{Alex_2020_area-metric-gravity}. We display the 16-dimensional solution for the Lagrangian up to second order, with 16 undetermined constants $k_{1},\dots,k_{16}$. For the solution of the third order, with 50 undetermined constants $k_{1},\dots,k_{50}$, we point to the aforementioned reference.

\begingroup
\allowdisplaybreaks
\begin{align}\label{reductions}
  e_{1} = {} &  k_{1} \nonumber \\
  e_{2} = {} &  k_{2} \nonumber \\
  e_{3} = {} & -2 k_{1} -\frac{2}{3} k_{2} \nonumber \\
  e_{4} = {} & 4 k_{1} + \frac{1}{3} k_{2} \nonumber \\
  e_{5} = {} &  k_{3} \nonumber \\
  e_{6} = {} & -3 k_{1} -\frac{1}{2} k_{2} -3 k_{3} \nonumber \\
  e_{7} = {} &  k_{4} \nonumber \\
  e_{8} = {} &  k_{5} \nonumber \\
  e_{9} = {} &  k_{6} \nonumber \\
  e_{10} = {} &  k_{7} \nonumber \\
  e_{11} = {} &  k_{8} \nonumber \\
  e_{12} = {} & \frac{1}{2} k_{6} + \frac{5}{8} k_{7} \nonumber \\
  e_{13} = {} & -\frac{16}{3} k_{4} + 16 k_{5} -\frac{7}{3} k_{6} -\frac{5}{12} k_{7} + \frac{4}{3} k_{8} \nonumber \\
  e_{14} = {} & -\frac{8}{3} k_{4} + 8 k_{5} -\frac{13}{6} k_{6} -\frac{11}{24} k_{7} + \frac{2}{3} k_{8} \nonumber \\
  e_{15} = {} &  k_{4} -\frac{1}{8} k_{6} -\frac{23}{32} k_{7} -\frac{1}{2} k_{8} \nonumber \\
  e_{16} = {} &  k_{9} \nonumber \\
  e_{17} = {} &  k_{10} \nonumber \\
  e_{18} = {} & \frac{3}{2} k_{4} + \frac{3}{4} k_{6} -\frac{3}{16} k_{7} + 3 k_{9} \nonumber \\
  e_{19} = {} & \frac{1}{2} k_{4} + \frac{1}{4} k_{6} -\frac{1}{16} k_{7} +  k_{9} \nonumber \\
  e_{20} = {} & -\frac{1}{4} k_{4} -\frac{1}{8} k_{6} + \frac{1}{32} k_{7} -\frac{1}{2} k_{9} \nonumber \\
  e_{21} = {} &  k_{4} -3 k_{5} + \frac{1}{4} k_{6} -\frac{3}{16} k_{7} -\frac{1}{2} k_{8} +  k_{9} -3 k_{10} \nonumber \\
  e_{22} = {} &  k_{11} \nonumber \\
  e_{23} = {} &  k_{12} \nonumber \\
  e_{24} = {} &  k_{13} \nonumber \\
  e_{25} = {} &  k_{14} \nonumber \\
  e_{26} = {} &  k_{6} + \frac{3}{4} k_{7} - k_{14} \nonumber \\
  e_{27} = {} & - k_{4} + \frac{1}{2} k_{7} \nonumber \\
  e_{28} = {} & \frac{5}{3} k_{4} + \frac{5}{12} k_{6} -\frac{25}{48} k_{7} -2 k_{11} - k_{12} -\frac{2}{3} k_{13} -\frac{1}{4} k_{14} \nonumber \\
  e_{29} = {} &  k_{6} + \frac{3}{4} k_{7} - k_{14} \nonumber \\
  e_{30} = {} & -\frac{4}{3} k_{4} -\frac{5}{6} k_{6} + \frac{1}{24} k_{7} + 4 k_{11} + 2 k_{12} + \frac{1}{3} k_{13} + \frac{1}{2} k_{14} \nonumber \\
  e_{31} = {} &  k_{15} \nonumber \\
  e_{32} = {} &  k_{16} \nonumber \\
  e_{33} = {} &  k_{4} -\frac{1}{2} k_{7} -3 k_{11} -\frac{1}{2} k_{13} -6 k_{15} \nonumber \\
  e_{34} = {} & \frac{1}{2} k_{6} + \frac{3}{8} k_{7} -\frac{3}{2} k_{12} -\frac{1}{2} k_{14} -3 k_{16} \nonumber \\
  e_{35} = {} & -2 k_{4} - k_{6} + \frac{1}{4} k_{7} \nonumber \\
  e_{36} = {} & - k_{4} + \frac{1}{2} k_{7} -\frac{3}{2} k_{12} -\frac{1}{2} k_{14} -3 k_{16} \nonumber \\
  e_{37} = {} & \frac{1}{12} k_{4} + \frac{1}{12} k_{6} + \frac{1}{48} k_{7} -\frac{1}{8} k_{12} -\frac{1}{24} k_{14} +  k_{15} + \frac{1}{4} k_{16} \nonumber \\
  e_{38} = {} & -2 k_{4} +  k_{7} \nonumber \\
  e_{39} = {} & -2 k_{6} -\frac{3}{2} k_{7} \nonumber \\
  e_{40} = {} &  k_{4} + \frac{1}{2} k_{6} -\frac{1}{8} k_{7}
\end{align}
\endgroup

\section{Linearized field equations}
\label{linear-eom}
\subsection{Unconstrained equations}
A three-plus-one split of the Lagrangian built from the ansätze above and subsequent variations w.r.t.~the spatial perturbation variables introduced in Sec.~\ref{three_plus_one_split} yields the perturbative field equations for area metric gravity. We further split these equations into traceless tensor, vector, and scalar parts and apply the gauge condition from Sec.~\ref{three_plus_one_split}. This yields the linearized field equations for the scalar-trace and scalar-tracefree modes
\begin{equation}
  \label{scalar-equations-appendix}
  \begin{aligned}
    \left\lbrack\frac{\delta L}{\delta u^{\alpha\beta}}\right\rbrack^\text{S-TF} = {} & \Delta_{\alpha\beta} \left\lbrack s_1 A -\frac{s_1}{4} \tilde U + s_3 \tilde V + s_4 \ddot V - \frac{s_4}{3} \Delta V + s_6 \ddot W - \frac{s_6}{3} \Delta W\right\rbrack, \\
    \left\lbrack\frac{\delta L}{\delta v^{\alpha\beta}}\right\rbrack^\text{S-TF} = {} & \Delta_{\alpha\beta} \left\lbrack (s_1 + 4s_4) A + (\frac{s_1}{4} + s_4) \tilde U + (\frac{3s_1}{4} + 3 s_4) \tilde V \right. \\ & + \left. s_{11} \ddot V - (\frac{s_1}{3} + \frac{4s_4}{3} + s_{11}) \Delta V + s_{13} V + s_{14} \Box W + s_{16} W \right\rbrack, \\
    \left\lbrack\frac{\delta L}{\delta w^{\alpha\beta}}\right\rbrack^\text{S-TF} = {} & \Delta_{\alpha\beta} \left\lbrack 4s_6 A + s_6 \tilde U + 3 s_6 \tilde V \right. \\ & + \left. (-s_6 + s_{14}) \ddot V - (\frac{s_6}{3} + s_{14}) \Delta V + s_{16} V - (\frac{s_1}{4} + s_4 + s_{11}) \Box W - s_{13} W \right\rbrack, \\
    \left\lbrack\frac{\delta L}{\delta u^{\alpha\beta}}\right\rbrack^\text{S-TR} = {} & \gamma_{\alpha\beta} \left\lbrack -\frac{2s_1}{3} \Delta A -\frac{s_1}{2} \ddot{\tilde U} + \frac{s_1}{6} \Delta\tilde U + (-\frac{3s_1}{4} + s_3) \ddot{\tilde V} -\frac{2s_3}{3} \Delta\tilde V \right. \\ & \left. + \frac{s_1}{3} \Delta\ddot V + \frac{2s_4}{9} \Delta\Delta V + \frac{2s_6}{9} \Delta\Delta W\right\rbrack, \\
    \left\lbrack\frac{\delta L}{\delta v^{\alpha\beta}}\right\rbrack^\text{S-TR} = {} & \gamma_{\alpha\beta} \left\lbrack (-s_1 + \frac{4s_3}{3}) \Delta A + (-\frac{3s_1}{4} + s_3) \ddot{\tilde U} - \frac{2s_3}{3} \Delta\tilde U \right. \\ & \left. + s_{37} \ddot{\tilde V} - (\frac{3s_1}{2} - 2s_3 + s_{37}) \Delta\tilde V + s_{39} \tilde V \right. \\ & \left. + (\frac{s_1}{2} - \frac{2s_3}{3}) \Delta\ddot V + (\frac{s_1}{6} + \frac{2s_3}{9} + \frac{2s_4}{3}) \Delta\Delta V + \frac{2s_6}{3} \Delta\Delta W\right\rbrack, \\
    \left\lbrack\frac{\delta L}{\delta b^{\alpha}}\right\rbrack^\text{S} = {} & \partial_\alpha \partial_t \left\lbrack -2s_1\tilde U + (-3s_1 + 4s_3) \tilde V + (\frac{4s_1}{3} + \frac{8s_4}{3}) \Delta V + \frac{8s_6}{3} \Delta W\right\rbrack, \\
    \frac{\delta L}{\delta A} = {} & -2s_1 \Delta\tilde U + (-3s_1 + 4s_3) \Delta\tilde V + (\frac{4s_1}{3} + \frac{8s_4}{3}) \Delta\Delta V + \frac{8s_6}{3} \Delta\Delta W,
  \end{aligned}
\end{equation}
which depend on 10 independent combinations $s_i$ of the 16 undetermined constants $k_i$,
\begingroup
\allowdisplaybreaks
\begin{align}
    s_{1} = {} &  2 k_{6} + \frac{3}{2} k_{7}, \nonumber \\
    s_{3} = {} &  \frac{3}{2} k_{6} + \frac{9}{8} k_{7} - 6 k_{12} - 2 k_{14}, \nonumber \\
    s_{4} = {} &  -\frac{1}{2} k_{6} - \frac{3}{8} k_{7} - \frac{1}{2} k_{14}, \nonumber \\
    s_{6} = {} &  k_{6} + \frac{3}{4} k_{7} - 3 k_{12} - k_{14} - 6 k_{16}, \nonumber \\
    s_{11} = {} &  \frac{1}{2} k_{6} + \frac{11}{8} k_{7} + 2 k_{8} - 2 k_{13} - \frac{1}{2} k_{14}, \nonumber \\
    s_{13} = {} &  -2 k_{2}, \nonumber \\
    s_{14} = {} &  -2 k_{4} + 24 k_{5} - k_{6} - \frac{3}{4} k_{7} + 4 k_{8} - 12 k_{9} + 24 k_{10} - 24 k_{11} - 6 k_{12} - 4 k_{13} \nonumber \\ & - 2 k_{14} - 48 k_{15} - 12 k_{16}, \nonumber \\
    s_{16} = {} &  -24 k_{1} - 4 k_{2} - 24 k_{3}, \nonumber \\
    s_{37} = {} &  -24 k_{5} + 2 k_{6} + \frac{5}{2} k_{7} - 4 k_{8} + 24 k_{11} - 12 k_{12} + 4 k_{13} - 4 k_{14}, \nonumber \\
    s_{39} = {} &  24 k_{1} + 4 k_{2}.
\end{align}
\endgroup
A subset of 7 constants $s_i$ governs the field equations for the vector modes,
\begin{equation}
  \begin{aligned}
    \left\lbrack\frac{\delta L}{\delta u^{\alpha\beta}}\right\rbrack^\text{V} = {} & \partial_t \partial_{(\alpha} \left\lbrack s_1 B_{\beta)} - 2s_4 \dot U_{\beta)} - 2s_6 \epsilon_{\beta)}^{\hphantom{\beta)}\mu\nu} U_{\mu,\nu} + 2s_6 \dot W_{\beta)} + (-\frac{s_1}{2} - 2s_4) \epsilon_{\beta)}^{\hphantom{\beta)}\mu\nu} W_{\mu,\nu} \right\rbrack, \\
    \left\lbrack\frac{\delta L}{\delta v^{\alpha\beta}}\right\rbrack^\text{V} = {} & \partial_{(\alpha} \left\lbrack (-s_1 - 4s_4) \dot B_{\beta)} + 4s_6 \epsilon_{\beta)}^{\hphantom{\beta)}\mu\nu} B_{\mu,\nu} \right. \\ & + (s_1 + 4s_4 + 2s_{11}) \ddot U_{\beta)} + (-\frac{3s_1}{2} - 6s_4 - 2s_{11}) \Delta U_{\beta)} + 2s_6 \epsilon_{\beta)}^{\hphantom{\beta)}\mu\nu} \dot U_{\mu,\nu} + 2s_{13} U_{\beta)} \\ & \left. + 2 s_{14} \Box W_{\beta)} + 2 s_{16} W_{\beta)} \right\rbrack, \\
    \left\lbrack\frac{\delta L}{\delta w^{\alpha\beta}}\right\rbrack^\text{V} = {} & \partial_{(\alpha} \left\lbrack 4s_6 \dot B_{\beta)} + (s_1 + 4s_4) \epsilon_{\beta)}^{\hphantom{\beta)}\mu\nu} B_{\mu,\nu} \right. \\ & + (2s_6 + 2 s_{14}) \ddot U_{\beta)} - 2 s_{14} \Delta U_{\beta)} + (\frac{s_1}{2} + 2 s_4) \epsilon_{\beta)}^{\hphantom{\beta)}\mu\nu} \dot U_{\mu,\nu} + 2 s_{16} U_{\beta)} \\ & \left. + (-\frac{3s_1}{2} - 6s_4 - 2s_{11}) \Box W_{\beta)} - 2s_{13} W_{\beta)} \right\rbrack, \\
    \left\lbrack\frac{\delta L}{\delta b^\alpha}\right\rbrack^\text{V} = {} & \Delta\left\lbrack 2s_1 B_{\alpha} - 4s_4 \dot U_{\alpha} - 4s_6 \epsilon_{\alpha}^{\hphantom{\alpha}\mu\nu} U_{\mu,\nu} + 4s_6 \dot W_{\alpha} + (-s_1 - 4s_4) \epsilon_{\alpha}^{\hphantom{\alpha}\mu\nu} W_{\mu,\nu} \right\rbrack,
  \end{aligned}
\end{equation}
as well as the traceless tensor modes
\begin{equation}
  \begin{aligned}
    \left\lbrack\frac{\delta L}{\delta u^{\alpha\beta}}\right\rbrack^\text{TT} = {} & \frac{s_1}{4} \Box U_{\alpha\beta}  \\ &  + (\frac{s_1}{4} + s_4) \ddot V_{\alpha\beta} + (\frac{s_1}{4} + s_4) \Delta V_{\alpha\beta} - 2s_6 \epsilon_{(\alpha}^{\hphantom{(\alpha}\mu\nu} \dot V_{\beta)\mu,\nu} \\ & + s_6 \ddot W_{\alpha\beta} + s_6 \Delta W_{\alpha\beta} + (\frac{s_1}{2} + 2 s_4) \epsilon_{(\alpha}^{\hphantom{(\alpha}\mu\nu} \dot W_{\beta)\mu,\nu}, \\
    \left\lbrack\frac{\delta L}{\delta v^{\alpha\beta}}\right\rbrack^\text{TT} = {} & (\frac{s_1}{4} + s_4) \ddot U_{\alpha\beta} + (\frac{s_1}{4} + s_4) \Delta U_{\alpha\beta} + 2s_6 \epsilon_{(\alpha}^{\hphantom{(\alpha}\mu\nu} \dot U_{\beta)\mu,\nu}  \\ &  + (\frac{s_1}{4} + s_4 + s_{11}) \Box V_{\alpha\beta} + s_{13} V_{\alpha\beta} + s_{14} \Box W_{\alpha\beta} + s_{16} W_{\alpha\beta}, \\
    \left\lbrack\frac{\delta L}{\delta w^{\alpha\beta}}\right\rbrack^\text{TT} = {} & s_6 \ddot U_{\alpha\beta} + s_6 \Delta U_{\alpha\beta} - (\frac{s_1}{2} + 2 s_4) \epsilon_{(\alpha}^{\hphantom{(\alpha}\mu\nu} \dot U_{\beta)\mu,\nu}  \\ &  + s_{14} \Box V_{\alpha\beta} + s_{16} V_{\alpha\beta} - (\frac{s_1}{4} + s_4 + s_{11}) \Box W_{\alpha\beta} - s_{13} W_{\alpha\beta}.
  \end{aligned}
\end{equation}
Note that the Noether identities \cite{Alex_2020_PhD} $0 = \partial_t\frac{\delta L}{\delta A} - \partial_\alpha\frac{\delta L}{\delta b_\alpha}$ and $0 = \partial_t\frac{\delta L}{\delta b^\alpha} - 4 \partial_{\beta}\frac{\delta L}{\delta u_{\alpha\beta}}$ are easily verified.

\subsection{Constrained equations}
As discussed in Sec.~\ref{three_plus_one_split}, the field equations exhibit behavior we deem of unphysical phenomenology for a theory which shall only introduce refinements to Einstein gravity. First, the scalar equations (\ref{scalar-equations-appendix}) yield, among short-ranging Yukawa corrections, long-ranging Coulomb corrections to the linearized Schwarzschild solution, \emph{except for}
\begin{equation}
  s_1 + 4 s_4 = 0 \quad\text{and}\quad s_6 = 0.
\end{equation}
Constraining the theory to this sector yields the scalar field equations
\begin{equation}\label{scalar-eqns-reduced}
  \begin{aligned}
    \left\lbrack\frac{\delta L}{\delta u^{\alpha\beta}}\right\rbrack^\text{S-TF} = {} & \Delta_{\alpha\beta} \left\lbrack s_1 A -\frac{s_1}{4} \tilde U + s_3 \tilde V - \frac{s_1}{4} \ddot V + \frac{s_1}{12} \Delta V \right\rbrack, \\
    \left\lbrack\frac{\delta L}{\delta v^{\alpha\beta}}\right\rbrack^\text{S-TF} = {} & \Delta_{\alpha\beta} \left\lbrack s_{11} \Box V + s_{13} V + s_{14} \Box W + s_{16} W \right\rbrack, \\
    \left\lbrack\frac{\delta L}{\delta w^{\alpha\beta}}\right\rbrack^\text{S-TF} = {} & \Delta_{\alpha\beta} \left\lbrack s_{14} \Box V + s_{16} V - s_{11} \Box W - s_{13} W \right\rbrack, \\
    \left\lbrack\frac{\delta L}{\delta u^{\alpha\beta}}\right\rbrack^\text{S-TR} = {} & \gamma_{\alpha\beta} \left\lbrack -\frac{2s_1}{3} \Delta A -\frac{s_1}{2} \ddot{\tilde U} + \frac{s_1}{6} \Delta\tilde U + (-\frac{3s_1}{4} + s_3) \ddot{\tilde V} -\frac{2s_3}{3} \Delta\tilde V \right. \\ & \left. + \frac{s_1}{3} \Delta\ddot V - \frac{s_1}{18} \Delta\Delta V \right\rbrack, \\
    \left\lbrack\frac{\delta L}{\delta v^{\alpha\beta}}\right\rbrack^\text{S-TR} = {} & \gamma_{\alpha\beta} \left\lbrack (-s_1 + \frac{4s_3}{3}) \Delta A + (-\frac{3s_1}{4} + s_3) \ddot{\tilde U} - \frac{2s_3}{3} \Delta\tilde U \right. \\ & \left. + s_{37} \ddot{\tilde V} - (\frac{3s_1}{2} - 2s_3 + s_{37}) \Delta\tilde V + s_{39} \tilde V \right. \\ & \left. + (\frac{s_1}{2} - \frac{2s_3}{3}) \Delta\ddot V + \frac{2s_3}{9} \Delta\Delta V\right\rbrack, \\
    \left\lbrack\frac{\delta L}{\delta b^{\alpha}}\right\rbrack^\text{S} = {} & \partial_\alpha \partial_t \left\lbrack -2s_1\tilde U + (-3s_1 + 4s_3) \tilde V + \frac{2s_1}{3} \Delta V\right\rbrack, \\
    \frac{\delta L}{\delta A} = {} & -2s_1 \Delta\tilde U + (-3s_1 + 4s_3) \Delta\tilde V + \frac{2s_1}{3} \Delta\Delta V,
  \end{aligned}
\end{equation}
the vector field equations
\begin{equation}
  \begin{aligned}
    \left\lbrack\frac{\delta L}{\delta u^{\alpha\beta}}\right\rbrack^\text{V} = {} & \frac{s_1}{2} \partial_t \partial_{(\alpha} \left\lbrack 2 B_{\beta)} + \dot U_{\beta)}\right\rbrack, \\
    \left\lbrack\frac{\delta L}{\delta v^{\alpha\beta}}\right\rbrack^\text{V} = {} & 2 \partial_{(\alpha} \left\lbrack  s_{11} \Box U_{\beta)} + s_{13} U_{\beta)} +  s_{14} \Box W_{\beta)} +  s_{16} W_{\beta)} \right\rbrack, \\
    \left\lbrack\frac{\delta L}{\delta w^{\alpha\beta}}\right\rbrack^\text{V} = {} & 2 \partial_{(\alpha} \left\lbrack s_{14} \Box U_{\beta)} + s_{16} U_{\beta)} - s_{11} \Box W_{\beta)} - s_{13} W_{\beta)} \right\rbrack, \\
    \left\lbrack\frac{\delta L}{\delta b^\alpha}\right\rbrack^\text{V} = {} & s_1 \Delta\left\lbrack 2 B_{\alpha} + \dot U_{\alpha} \right\rbrack,
  \end{aligned}
\end{equation}
and the traceless tensor field equations
\begin{equation}\label{tt-eqns-reduced}
  \begin{aligned}
    \left\lbrack\frac{\delta L}{\delta u^{\alpha\beta}}\right\rbrack^\text{TT} = {} & \frac{s_1}{4} \Box U_{\alpha\beta}, \\
    \left\lbrack\frac{\delta L}{\delta v^{\alpha\beta}}\right\rbrack^\text{TT} = {} & s_{11} \Box V_{\alpha\beta} + s_{13} V_{\alpha\beta} + s_{14} \Box W_{\alpha\beta} + s_{16} W_{\alpha\beta}, \\
    \left\lbrack\frac{\delta L}{\delta w^{\alpha\beta}}\right\rbrack^\text{TT} = {} & s_{14} \Box V_{\alpha\beta} + s_{16} V_{\alpha\beta} - s_{11} \Box W_{\alpha\beta} - s_{13} W_{\alpha\beta}.
  \end{aligned}
\end{equation}
As explained in Sec.~\ref{three_plus_one_split}, the coupled wave equations of the kind
\begin{equation}
  \begin{aligned}
    u = {} & s_{11} \Box\varphi + s_{13} \varphi + s_{14} \Box\psi + s_{16} \psi, \\
    v = {} & s_{14} \Box\varphi + s_{16} \varphi - s_{11} \Box\psi - s_{13} \psi
  \end{aligned}
\end{equation}
lead to diverging behavior of solutions, \emph{unless}
\begin{equation}
   s_{13} s_{14} - s_{11} s_{16} = 0.
\end{equation}
We thus enforce $s_{16} = s_{13} s_{14}/s_{11}$ and arrive at a phenomenologically relevant subsector of the unconstrained linearized theory with only 7 independent gravitational constants left.

Of these 7 constants, 5 combinations appear in the linearized Schwarzschild solution (\ref{scalar-solution}) and the gravitational-wave solutions (\ref{area-metric-solution}) and (\ref{massive-radiating}), two constants which play the role of \emph{masses} in wave equations or screened Poisson equations,
\begin{equation}
  \mu^2 = \frac{8s_1s_{39}}{9s_1^2-24s_1s_3+8s_1s_{37}+16s_3^2} \quad\text{and}\quad \nu^2 = \frac{s_{11} s_{13} + s_{14} s_{16}}{s_{11}^2 + s_{14}^2},
\end{equation}
and three other constants
\begin{equation}
  \begin{aligned}
    \alpha  = {} & \frac{1}{2s_1}, \\
    \beta   = {} & \frac{(3s_1+4s_3)^2}{6s_1(9s_1^2-24s_1s_3+8s_1s_{37}+16s_3^2)}, \\
    \gamma  = {} & \frac{-8(3s_1+4s_3)}{6(9s_1^2-24s_1s_3+8s_1s_{37}+16s_3^2)}.
  \end{aligned}
\end{equation}

\textsc{Cadabra}, \textsc{Mathematica}, and \textsc{Maple} code assisting calculations in this section is publicly available at \cite{Alex_2020_area-metric-gravity}.

\end{document}